\documentclass[aps,prd,11pt,reprint,nofootinbib,showkeys,unsortedaddress]{revtex4-2}
\usepackage{amsmath,amssymb,amsthm,mathrsfs,bbm}
\usepackage{latexsym,amscd,amsbsy,amsfonts,dsfont}
\usepackage{graphicx}
\usepackage[utf8]{inputenc}
\usepackage{subfigure}  
\usepackage{float}
\usepackage{slashed}
\usepackage{cancel}
\usepackage{multirow}
\usepackage{soul}
\usepackage{physics}
\usepackage{comment}
\usepackage[usenames,dvipsnames]{xcolor}
\usepackage[
      colorlinks=true,
      linktocpage=true,
      breaklinks=true,
      linkcolor=Aquamarine,
      urlcolor=Purple,
      filecolor=black,
      citecolor=Purple,
      menucolor=black,
      pdfstartview=FitV,
      bookmarksopen=true
      ]{hyperref}

\makeindex

\begin{document}

\title{
Analogue gravity simulation of superpositions of spacetimes
}

\author{Carlos Barcel\'o}
\email{carlos@iaa.es}
\affiliation{Instituto de Astrof\'{\i}sica de Andaluc\'{\i}a (IAA-CSIC), Glorieta de la Astronom\'{\i}a, 18008 Granada, Spain}
\author{Luis J. Garay}
\email{luisj.garay@ucm.es}
\affiliation{Departamento de F\'{\i}sica Te\'orica and IPARCOS, \\
Universidad Complutense de Madrid, 28040 Madrid, España}
\author{Gerardo Garc\'ia-Moreno}
\email{ggarcia@iaa.es}
\affiliation{Instituto de Astrof\'{\i}sica de Andaluc\'{\i}a (IAA-CSIC), Glorieta de la Astronom\'{\i}a, 18008 Granada, Spain}

\begin{abstract}
{ Taking the principles of quantum mechanics as they stand and applying them to gravity, leads to the conclusion that one might be able to generate superpositions of spacetimes, at least formally. We analyze such a possibility from an analogue gravity perspective. We present an analogue toy model consisting of a Bose-Einstein condensate in a double-well potential and identify the states that could potentially be interpreted as superposition of effective spacetimes. These states are unstable and the source of instability from a microscopic point of view can be related to the absence of a well-defined causal structure in the effective geometric description. We explore the consequences of these instabilities and argue that they resonate with Penrose's ideas about the decay that superpositions of states with sufficiently different gravitational fields associated should experience. }
\end{abstract}


\maketitle

{

\hypersetup{citecolor=black, linkcolor=black,urlcolor=black}

\tableofcontents

}

\maketitle

\section{Introduction}
\label{Sec:Introduction}

The search for a theory of quantum gravity has been one of the main driving forces in theoretical physics over the last century. Many approaches exist toward building a successful theory of quantum gravity, among them string theory~\cite{Green1987a,Green1987b} and canonical quantum gravity~\cite{Wheeler1968} in its modern formulation in terms of loop quantization~\cite{Thiemann2007} are the leading approaches; although there are many other approaches as Asymptotic Safety~\cite{Percacci2017,Bonanno2020}, Causal Dynamical Triangulations~\cite{Loll2019}, or Causal Sets~\cite{Surya2019}. In any case, it is fair to say that none of them is regarded as being completely satisfactory. The laws of quantum physics are maintained in all of these theories and one simply applies different quantization schemes to the geometric degrees of freedom, or regards them as emergent and applies the standard quantum rules to other microscopic degrees of freedom. However, one remarkable property shared by many of these approaches is that it is possible to generate superpositions of almost classical states, i.e. states that we can almost describe as approximate smooth geometries at low energies. This very same notion of superposing semiclassical spacetimes has also a counterpart with the electromagnetic field, where even Schr\"odinger-cat-like states can be generated~\cite{Brune1992}. 

Our aim in this work is to dig into the physical meaning of superposing spacetimes, trying to simulate them within an analogue gravity framework~\cite{Barcelo2011}. The analogue gravity program intends to shed new light into gravitational physics by reproducing some of its behaviours in laboratory systems. Currently, analogue gravity is just able to reproduce kinematical properties of gravitational fields, not dynamical properties. In that sense, since our understanding of kinematical aspects of gravity is based on the theory of General Relativity, in which spacetime is described as a curved geometry~\cite{Wald1984}, we try to to simulate the propagation of signals on top of effectively curved geometries. Among the analogue gravity systems at our disposal, fluid systems and, in particular, quantum fluids based on Bose-Einstein Condensates (BECs) as their substratum, have proved to be very useful both at the theoretical and at the experimental level~\cite{Garay1999,Steinhauer2016}. This analogy relies on the long-wavelength regime of acoustic perturbations: such perturbations behave as if they were a scalar field moving in an effectively curved spacetime.

Whereas the standard analogy for a classical spacetime is well understood (see~\cite{Visser1997} for a systematic presentation) our aim here is to explore whether such analogy can be extended to the case of having superpositions of semiclassical spacetimes. The effective spacetime metric depends on the characteristics of the macroscopic wave function of a BEC which is a fully quantum mechanical object. Therefore, within this framework it appears in principle possible to engineer situations such that there exist acoustic waves perceiving a superposition of two different spacetimes, i.e. two different causalities. Whether this can be done or not is the main inquiry that we address in this work. For that purpose, we study a toy model consisting in a BEC in a double-well potential and look for the states that resemble a superposition of spacetimes. 

Given the insatisfactory situation regarding our current theories of quantum gravity, there has been a relatively small but steady trend pointing out that maybe one should not quantize gravity but \textit{gravitizate} quantum mechanics, in Penrose's words~\cite{Penrose2014}. One of the striking phenomena that we face when assuming that the laws of quantum physics are valid at all scales and for arbitrary objects is that we observe no phenomenology associated with quantum superpositions of macroscopic states although they are allowed within the conceptual framework of quantum mechanics. The standard explanation relies on a process known as environmental decoherence~\cite{Zeh1970,Zurek1981,Zurek1982}. However, Penrose has heuristically argued that gravity could be at the heart of some more fundamental source of decoherence~\cite{Penrose1989,Penrose1992,Penrose1996,Penrose2014}. As we will see, our results in the analogue gravity model strongly resonate with Penrose's ideas, although some differences also appear. Essentially we find that whenever some aspects of our analogue quantum gravity involve pushing the causal behaviour of General Relativity to its limits, either by superposing causalities as we do here, or by engineering chronologically pathological spacetimes as in~\cite{Barcelo2022}, we find that there exist kinematical limitations toward building such states. However, the dynamical mechanism leading to their destabilization when we try to build them is highly dependent on the particular dynamics of the analogue model.

Here is an outline of the article. In Section.~\ref{Section:Rev_Analogue_Gravity} we review the standard approach toward analogue gravity in which single spacetimes are simulated. Section~\ref{Section:Superp_Spacetimes} constitutes the core of the article. In subsection~\ref{Subsection:Condensates_Double_Well} we will review the main properties of the toy model that we will consider for bosons in a double well, the Bose-Hubbard model with two sites. In subsection~\ref{Subsection:Effective_SpacetimesE}, we will explain which are the main properties that this toy model needs to fullfill so that its sound wave excitations admit an effective description in terms of a Klein-Gordon equation propagating on top of a curved geometry. In subsection~\ref{Subsection:Stability}, we will explore the stability of the ground states of the Bose-Hubbard model depending on the possible values of its parameters. In subsection~\ref{Subsection:Superposing_Spacetimes}, we will study the implications of this stability analysis and discuss the constraints that we have to impose to the model in order to describe an effective spacetime. Furthermore, we will explore the possibility of building an effective superposition of spacetimes in this model. Finally, in subsection~\ref{Subsection:Same_Localization}, we will conclude by extending our discussion to the possibility of creating superposition of similar geometries in the same location but with different causalities. In Section.~\ref{Section:Penrose_Idea} we will discuss the similarities and differences between our findings in the analogue gravity system and Penrose's ideas on the superpositions of spacetimes. We will finish the paper by summarizing the main conclusions that can be drawn up from this work in Section~\ref{Section:Discussion}.

\textit{Notation and conventions.} We work in units in which $\hbar = c = 1$, unless explicitly stated. We will use the signature $(-,+,+,+)$ for the spacetime metric. Greek indices $(\mu, \nu, ...)$ will run from $0$ to $3$, representing spacetime indices, whereas latin indices $(i,j...)$ will run from $1$ to $3$ and represent spatial indices. Einstein's summation convention is used throughout the work unless otherwise stated.

\section{Analogue gravity: The Standard Picture}
\label{Section:Rev_Analogue_Gravity}

In general terms, for every barotropic and inviscid\footnote{Barotropic means that the equation of state of the fluid involves just a relation between its pressure $p$ and its density $\rho$, whereas inviscid means that the fluid viscosity is zero.} fluid whose flow is irrotational, it is possible to arrange the velocity potential $\phi$ describing sound waves on top of the fluid background to satisfy a massless Klein-Gordon equation~\cite{Visser1997}
\begin{equation}
    \frac{1}{\sqrt{-g}} \partial_{\mu} \left( \sqrt{-g} g^{\mu \nu} \partial_{\nu} \phi \right) = 0,
\end{equation}
where we have introduced the metric $g_{\mu \nu}$, its determinant $g=\text{det} g_{\mu\nu}$ and its inverse $g^{\mu \nu}$, defined in terms of the physical parameters as 
\begin{equation}
    g^{\mu\nu}=\frac{1}{\rho_0 c} \left(
    \begin{array}{c|c}
    -1 & -v^i \\
    \hline \\
    -v^j & c^2 \delta^{ij}-v^i v^j
    \end{array}
    \right),
\label{Eq:Acoustic_Inverse_Metric}
\end{equation}
with $c$ the local speed of sound, $v^i$ the velocity of the fluid, $\rho$ its density, and where we are writing everything in the $(t, \Vec{x}$) laboratory coordinates. This acoustic metric provides the sound-cones at each point of the spacetime.
One of the main properties of this acoustic geometry is that it is causally well-behaved. This was noticed in~\cite{Visser1997}, where it was shown that this metric automatically inherits the stable causality property~\cite{Hawking1973,Wald1984}
\begin{align}
    g^{\mu \nu} \nabla_{\mu} t \nabla_{\nu} t = - \frac{1}{\rho c} < 0. 
\end{align}
For a careful analysis on this point and a discussion on the possibility of building causally pathological acoustic geometries, see~\cite{Barcelo2022}.  

This will be the departure point of our discussion. Actually, up to this point, we have not made any comments on the nature of the fluid under consideration. It can be either an ordinary classical fluid or a quantum fluid. We will focus here on a particular kind of quantum fluid, a Bose-Einstein condensate, which is experimentally feasible to simulate spacetimes of interest~\cite{Garay2000,Steinhauer2016}. Actually, among the whole plethora of analogue systems that we have~\cite{Barcelo2011}, BECs are one of the best analogues since they are very clean and insensitive to external noise. 

From the perspective of this work the crucial characteristics of using a BEC as analogue system is that it is an essentially quantum system: it is close to the quantum vacuum state and in this sense very far from classicallity. The fluid analogy with a BEC starts from describing many-body system of bosons through a quantum field $\widehat{\Psi}$, whose dynamics is encoded in
\begin{align}
    i \frac{\partial}{\partial t} \widehat{\Psi} = \left(- \frac{1}{2 m} \Delta + V (\Vec{x}) + g \widehat{\Psi}^{\dagger} \widehat{\Psi} \right) \widehat{\Psi},
\end{align}
with $g$ the leading order interaction among bosons. At sufficiently low temperatures, these systems of weakly interacting bosons can develop condensation in the sense that a macroscopic number of particles populates the same state. To describe this condensation, it is customary to break the field into a macroscopic condensate and a fluctuation as $\widehat{\Psi} = \psi + \widehat{\delta \psi}$, with the condensation expressed as $\langle \widehat{\Psi} \rangle = \psi$. Using a Madelung representation and considering a long wavelength approximation one can describe the system as quantum acoustic perturbations moving in a curved background spacetime of the form~(\ref{Eq:Acoustic_Inverse_Metric}). The velocity of the background fluid is    
$\Vec{v} = \nabla \theta /m$ associated with the phase $\theta$ of $\psi$:
\begin{align}
    \psi = \sqrt{n_c}  e^{- i \theta}.
\end{align}
The density $\rho_0 = m n_c$ and $c_s$ represents the speed of the phonons in the medium 
\begin{align}
    c_s^2 = \frac{g n_c}{m}. 
\end{align}
As mentioned above, for this picture to be consistent, we must require that the wavelengths of the sound excitations are large. In fact, they have to be larger than the so-called healing length of the condensate $\xi = 1/ (m c_s)$, i.e. $\lambda \gg 2 \pi \xi$. This healing length is the characteristic length scale at which the hydrodynamic approximation breaks down and one is able to unveil the acoustic description of the fluid begins to break down.

Given this setup that we have introduced in which we have a quantum fluid, it is possible to think about generating superpositions of macroscopic states of the BEC. Using the language of analogue models in principle one could, for example, engineer a BEC to be in a superposition of two rather different acoustic black hole spacetimes: two black holes with different surface gravities. Another option is to engineer the BEC to be in two equivalent spacetimes, for example a single acoustic black hole, centered around different laboratory points. In a quantum gravity language, this could be taken also to be the geometries of two different coexistent universes. In the following we are going to discuss this last example by analyzing a BEC in a double-well and how the different possible ground states of the system depend on the values of the different parameters. Thus, our inquiry is the following: what does this superposition mean from an analogue gravity point of view? Is it possible to interpret this superposition as an effective superposition of spacetimes? And what is the fate of acoustic excitations propagating on top of it? The next section is deserved to analyze a toy model in which this questions can be sharply formulated.

\section{Analogue gravity: Superposing Spacetimes}
\label{Section:Superp_Spacetimes}

\subsection{Condensates in a double well}
\label{Subsection:Condensates_Double_Well}

Let us assume that we have scalar bosons in a symmetric double-well potential where we will label the wells by $i=1,2$. We will make the simplifying assumption for the moment that there is only one relevant state in each well and particles within a well have a contact interaction controlled by the parameter $U$ which can be either repulsive ($U>0$) or attractive ($U<0$). Also, we will assume that we have a term describing tunneling between the wells controlled by the parameter $t$ which is the amplitude of probability for tunneling between both wells (not to be confused with the time $t$ introduced in the previous section). The bigger the parameter $t$, the higher the potential barrier will be. The Hamiltonian describing these particular interactions is the Bose-Hubbard Hamiltonian~\cite{Lewenstein2012} with just ``two sites'' in the lattice
\begin{equation}
H = - t (a_1^{\dagger} a_2 + a_2^{\dagger} a_1 ) + \frac{U}{2}[n_1(n_1-1) + n_2 (n_2 -1) ],
\label{bosehubbard}
\end{equation}
where $a_i^{\dagger}, a_i$ represent the usual creation and annihilation operators that annihilate and create particles in the well $i$, with $n_i=a_i^{\dagger} a_i$ being the number of particles in the well $i$. Furthermore, we will work under the assumption of having a fixed number of particles $N = n_1 + n_2$. For the non-interacting Hamiltonian, i.e. Eq.~\eqref{bosehubbard} with only the tunneling term, the solutions are explicit and can be found via an ordinary Bogolyubov transformation. In the regime in which the tunneling is highly suppressed, just the interacting term survives. It is diagonal in the number basis and the solutions are also explicit since they simply correspond to the number basis. Notice that we are omitting the hats to denote the quantum operators from now on, to avoid a cumbersome notation.

We will now describe the main properties of the states which will be interesting for our purposes: the ground state of the free Hamiltonian and the ground states for the Hamiltonian with $t=0$, i.e., the ones with no hopping between the two wells, either with attractive $U<0$ or repulsive interactions $U>0$. They will serve as approximate ground states for the regimes of hopping domination $t\gg \abs{U}N$ and interaction domination $t \ll \abs{U}N $, respectively. The intermediate ground states between both regimes are not so useful for our purposes and can be found numerically. However, understanding the evolution of the system from one regime to the other as we adiabatically vary the parameters sheds some light onto the behaviour of the model and it can be well captured by the ansatz provided in~\cite{Mueller2006}. 

\begin{figure}
    \begin{center}
        \includegraphics[width = 0.3 \textwidth]{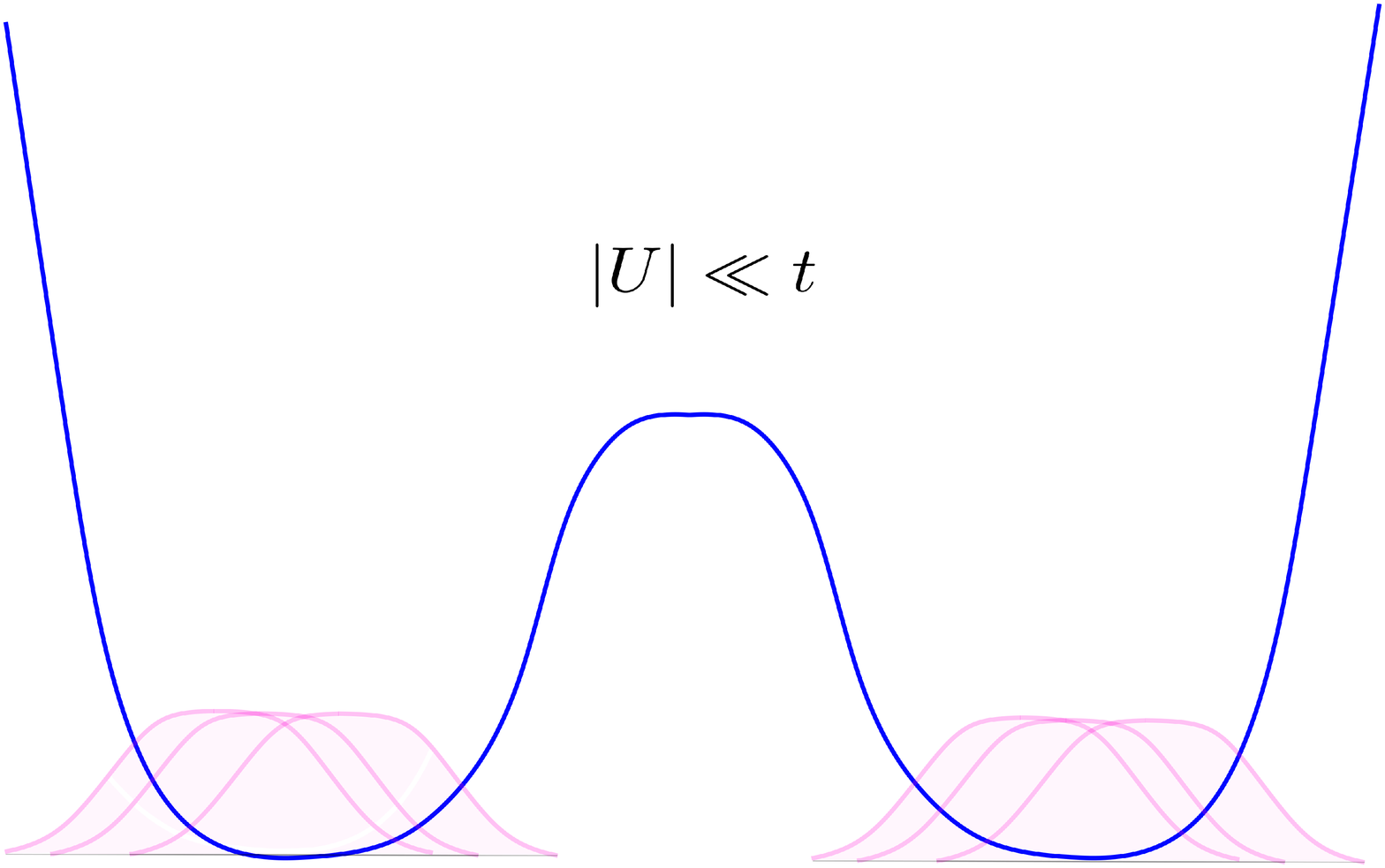}
        \includegraphics[width = 0.3 \textwidth]{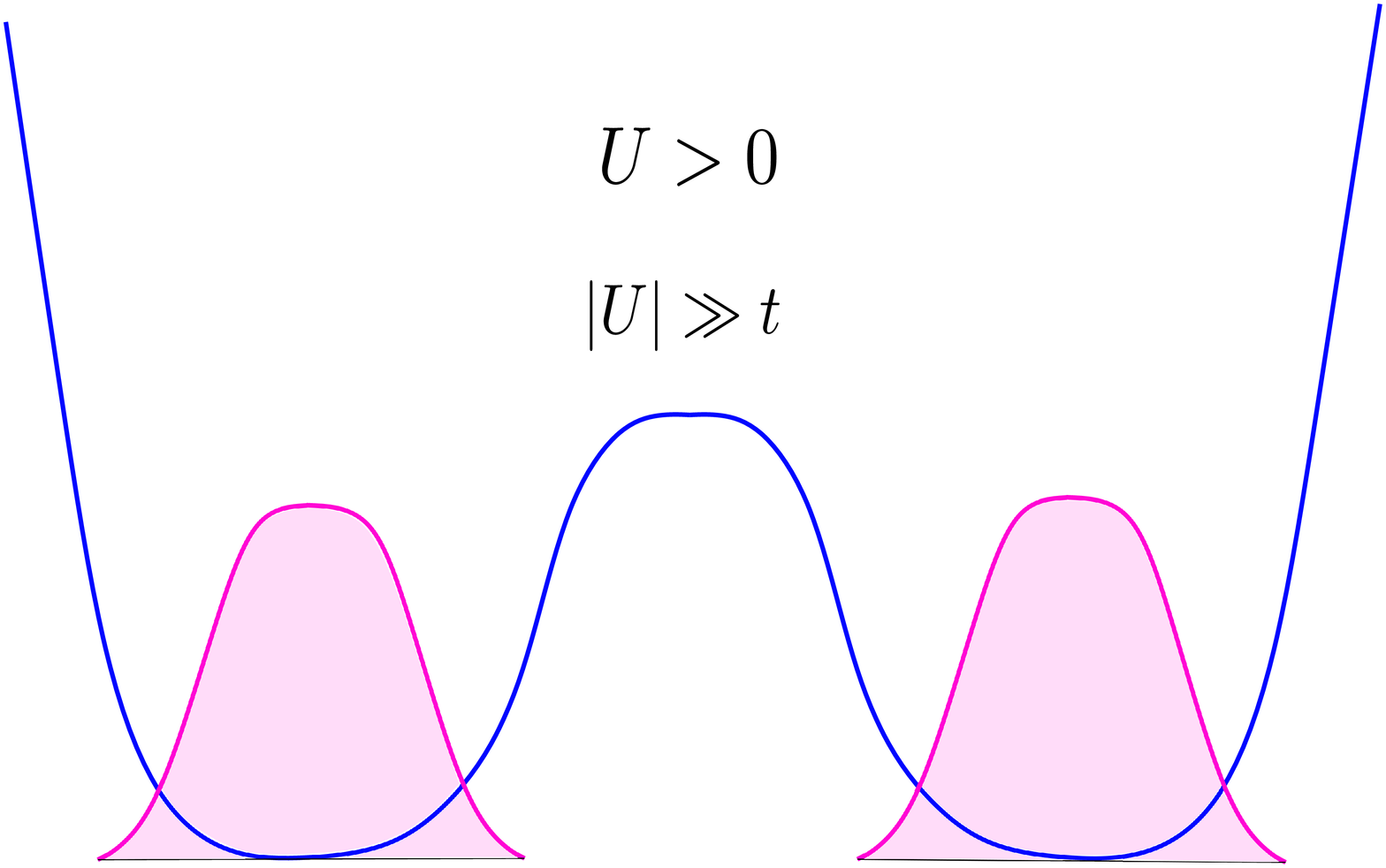}
        \includegraphics[width = 0.3 \textwidth]{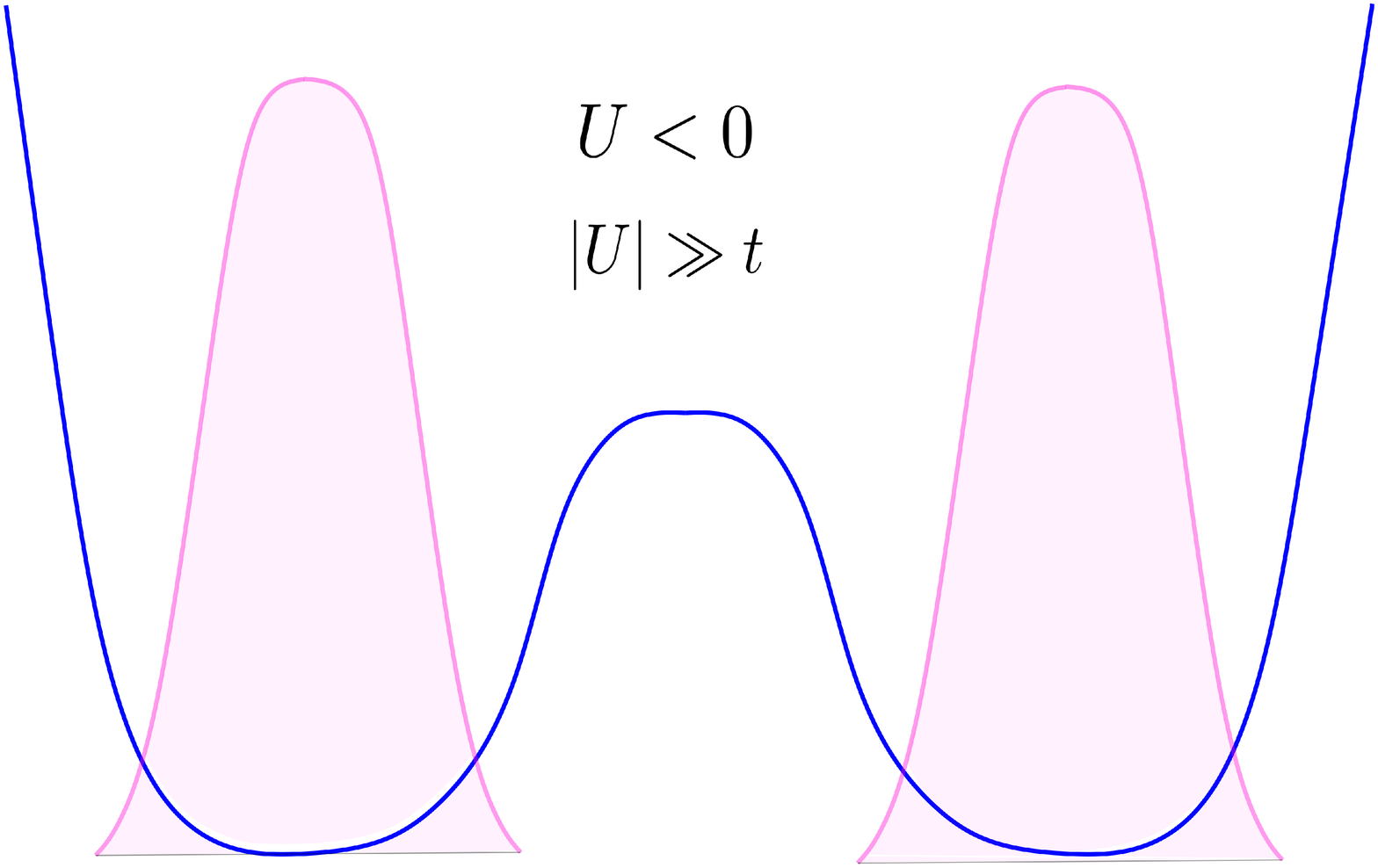}
        \caption{We pictorially represent the ground states corresponding to the three regimes of interest. They are depicted in the order in which they appear in the text (from top to the bottom): the coherent state, the Fock state and the cat state. The opacity in the figures above represents how localized the state is. The coherent state is a completely delocalized state with a non-vanishing projection onto all of the basis number states, the Fock state corresponds to two localized places at which the state exhibits a peak and the cat state corresponds to a delocalized state corresponding to all the particles being in one of the wells or the other.}
        \label{Fig:Ground_States}
    \end{center}
\end{figure}

Prior to discussing the ground states on the extreme regimes above, we need to introduce the concept of the single-particle reduced density matrix $\rho_{\textsc{sp}}$. Such density matrix is defined as 
\begin{equation}
\left( \rho_{\textsc{sp}} \right)_{ij} = \tr \left( \rho a^{\dagger}_{i} a_j  \right),
\end{equation}
where $\rho$ represents the density matrix of the system. According to Leggett's nomenclature~\cite{Leggett2006}, a condensate will be fragmented if its single-particle density matrix has two macroscopic eigenvalues, i.e., two eigenstates with eigenvalue of order $\order{N}$. Intuitively, this would correspond to a situation in which condensation occurs simultaneously in two different single-particle states. Armed with this tool, let us now analyze the three ground states of interest: the free case, the repulsive interacting case and the attractive interacting case. The corresponding states are pictorially represented in Fig.~\ref{Fig:Ground_States}

\textbf{Free condensate:} The free case can be explicitly solved via an ordinary Bogolyubov transformation, described by the following $SO(2)$ rotation within the space of operators $\{ a_1^{\dagger}, a_2^{\dagger} \}$: 
\begin{equation}
b_{1,2}^{\dagger} = \left(a_1^{\dagger} \pm a_2^{\dagger} \right)/\sqrt{2}.
\end{equation}
Thus, we find that for the single-particle subspace the eigenstates of the Hamiltonian are the symmetric and antisymmetric states with energies $-t$ and $t$, respectively. Thus, the ground state will be given by the fully symmetric state
\begin{equation}
\ket{C} = \frac{1}{\sqrt{2^N} N!} \left( a_1^{\dagger} + a_2^{\dagger}\right)^N \ket{0},
\end{equation}
because they do not interact among themselves. $\ket{0}$ represents the Fock vacuum, i.e., the state annihilated by $a_1$ and $a_2$. The $C$ stands for the coherent state, since it corresponds to $N$ bosons coherently delocalized among the two wells with equal probability of finding them in one of the wells. The single particle density matrix for this state reads
\begin{equation}
\rho_{\textsc{sp}} (C) = \frac{N}{2} \left( \begin{matrix} 1&1 \\ 1&1 \end{matrix} \right).
\end{equation}
This state has a single macroscopic eigenvalue $N$. According to Leggett's classification~\cite{Leggett2006}, it corresponds to a non-fragmented condensate since we just have one macroscopic eigenvalue of the single-particle density matrix. Physically, it corresponds to a condensation on the single-particle state which is approximately a superposition of the two Gaussians peaked around the center of each well in position space, describing the ground state of each well. Since the ground state $\ket{C}$ is a linear combination of number states $\ket{n_1,n_2} = a_1 ^{\dagger n_1} a_2 ^{\dagger n_2} \ket{0}/ \sqrt{n_1 ! n_2 !}$, the number of particles in each well has enormous fluctuations. We can compute them by writing the coherent state $\ket{C}$ in the number basis. For even $N$ (we consider such case for simplicity), we have 
\begin{equation}
\ket{C} = \sum_{\ell = -N/2}^{N/2} \Psi_{\ell}^{(0)} \ket{\ell},
\end{equation}
where $\ket{\ell} \equiv \ket{\frac{N}{2} + \ell, \frac{N}{2} - \ell}$ and 
\begin{equation}
\Psi_{\ell}^{(0)} = \left( \frac{N!}{2^N \left( \frac{N}{2} + \ell \right)! \left( \frac{N}{2} - \ell \right)!} \right)^{1/2} \approx \frac{e^{-\ell^2/N}}{(\pi N /2)^{1/4}}. 
\label{Gaussiandistrib}
\end{equation}
The fluctuations in the number of particles on each well are given by~\cite{Mueller2006}
\begin{equation}
\expval{\Delta n_i ^2}_C = \expval{(n_i - \expval{n_i}_C)^2}_C = N/4.
\end{equation}
with $i=1,2$. 

Let us move on to discuss the strongly interacting case, i.e. \mbox{$ t=0$}, \mbox{$U \neq 0$}. For this purpose, taking into account that the total number of particles is conserved \mbox{$n_1 + n_2 = N$}, we can rewrite the Hamiltonian as
\begin{equation}
H = \frac{U}{4} \left[ (n_1 - n_2)^2 + N^2 - 2N \right],
\end{equation}
which parametrizes the interaction by the difference in the number of particles on each well. 

\textbf{Strong repulsive interactions:} We will begin with the repulsive interactions $(U>0)$. Such case clearly favors the minimum difference in the number of particles between the two wells. Thus, for even $N$, the ground state for the repulsive interactions is clearly
\begin{equation}
\ket{F} = \frac{a_1^{\dagger N/2} a_2^{\dagger N/2}}{(N/2)!} \ket{0},
\label{fock_state}
\end{equation}
where the $F$ stands for the Fock state. Its single particle density matrix is
\begin{equation}
\rho_{\textsc{sp}} (F) = \frac{N}{2} \left( \begin{matrix} 1&0 \\ 0&1 \end{matrix} \right),
\label{matrixfock}
\end{equation}
which corresponds to a fragmented condensate since it has a macroscopic eigenvalue $N/2$ with multiplicity $2$. It corresponds to two uncorrelated condensates (notice that the state is a product state) having half the particles, each one located at one of the two wells. Such a result for the ground state in this limit is quite intuitive since having one more particle than the half of them on one of the two wells would imply paying a penalty in energy. The fluctuations in the number of particles on each well vanish identically
\begin{equation}
\expval{(\Delta n_i)^2}_F=0.
\end{equation}
Thus, in this case, we have a fragmented condensate with a well-defined number of particles on each well. 

\textbf{Strong attractive interactions:} For attractive interactions $U<0$, the situation is the opposite: the favored states are those containing a huge difference in the number of particles on each well. This means that the ground state is the subspace spanned by the vectors  $\{ \ket{N,0}, \ket{0,N} \}$. However, if we begin with the coherent state $\ket{C}$, the ground state of the free theory, and we turn on adiabatically the attractive interactions, the state that we will reach once $t$ becomes negligible will be a concrete superposition of both vectors (further details of how this state is reached will be provided at the end of the section). Actually, it corresponds to a Schr\"{o}dinger-cat like state
\begin{equation}
\ket{\textrm{cat}} = \frac{1}{\sqrt{2}} \left( \ket{N,0} + \ket{0,N} \right).
\label{cat_state}
\end{equation}
It is also fragmented since its single-particle denstiy matrix has also two eigenvalues 
\begin{equation}
\rho_{\textsc{sp}} (\textrm{cat}) = \frac{N}{2} \left( \begin{matrix} 1&0 \\ 0&1 \end{matrix} \right) ,
\end{equation}
being equal to $\rho_{\textsc{sp}} (F)$, the one that we had for the Fock state. Particle number fluctuations on each of the wells allow us to distinguish between both states though: in this case they do not vanish but
\begin{equation}
\expval{(\Delta n_i)^2}_{\textrm{cat}} = N^2/4.
\end{equation}
Thus, we have seen that although the Fock and cat states are fragmented according to the standard definition of Leggett (their one-particle density matrices have two macroscopic eigenvalues), they physically correspond to very different states. While the Fock state has a definite number of particles on each well, the cat state has an indefinite number of particles on each well; the latter corresponds to having the bosons delocalized. 

Until now, we have presented the three states that will be relevant for our purposes. They are the coherent state $\ket{C}$, which is the ground state of the free theory $t \neq 0, U =0$; the Fock state $\ket{F}$, which is the ground state of the strongly repulsive interacting theory $t=0, U >0$; and the cat state $\ket{\textrm{cat}}$, which is the ground state of the strongly attractive interacting theory $t=0, U<0$. The coherent state corresponds to a non-fragmented condensation while the Fock and cat states correspond to fragmented condensates. With respect to fluctuations in the number of particles on each well, the Fock state has no fluctuations on the number of particles on each well, while the coherent and cat states have enormous fluctuations on the number of particles on each well. These are the relevant features of these three states which will play a role in building our analogue models. We now provide additional details of how the model behaves for intermediate regimes.

\textbf{Intermediate interactions:} We will analyze what happens as we turn on the interactions adiabatically, i.e., we begin with the free Hamiltonian and assume we slowly turn on the interactions in such a way that the ground state of the free theory accommodates to the ground state of the interacting theory. Although numerically it is possible to obtain explicitly the wave function of the ground state for the interacting theory, as we advanced, we will take advantage of the ansatz introduced in~\cite{Mueller2006} and use their families of states as analytic states capturing the main properties of the ground states and being close to them. 

These ans\"{a}tze for the interacting theory will be based on the ground state wave function for the non-interacting theory, the $U=0$ case of the Hamiltonian (\ref{bosehubbard}). The starting point will be to write the ground state in terms of the number basis
\begin{equation}
\ket{\Psi} = \sum_{\ell=-N/2}^{N/2} \Psi_{\ell} \ket{\ell},
\end{equation}
and write the Schr\"{o}dinger equation for this system $ H \ket{\Psi} = E \ket{\Psi}$, which gives the following expression 
\begin{equation}
E \Psi_{\ell} = - t_{\ell + 1} \Psi_{\ell + 1} - t_{\ell} \Psi_{\ell -1} + U \ell^2 \Psi_{\ell},
\label{tightbinding}
\end{equation}
with 
\begin{equation}
t_{\ell} = t \sqrt{(N/2 + \ell)(N/2 - \ell +1 )}. 
\end{equation}
The problem is equivalent then to a one-dimensional tight-binding model in a harmonic potential with \mbox{non-uniform} tunneling matrix elements \cite{Mueller2006}. The non-uniformity is such that wave functions $\Psi_{\ell}$ with large amplitudes near $\ell \sim 0$ always have less energy than the ones that spread around different values of $\ell$. Actually, in the free case, the wave function was a narrow Gaussian centered at $\ell = 0$. 

With this expression for the eigenvalue problem, let us start by considering the repulsive case $(U>0)$. These interactions make the coherent state~\eqref{Gaussiandistrib} squeeze in an even narrower distribution. The family of states introduced in \cite{Mueller2006} for capturing this evolution from the coherent to the Fock state is
\begin{equation}
\Psi_{\ell}(\sigma) = \frac{e^{-\ell^2/\sigma^2}}{(\pi \sigma^2 / 2)^{1/4}}.
\label{familystates}
\end{equation}
As $\sigma^2$ varies from $ \sqrt{N}$ to small values, the initial coherent state $\ket{C}$ starts looking much more like the Fock state. Actually we can obtain a relation between $\sigma$ and the value of $U$ by taking a continuum limit on~\eqref{tightbinding}, see~\cite{Mueller2006} for details. In this limit, the equation reduces to that of a harmonic oscillator potential and we can obtain the value of $\sigma (U)$ since in that case the problem is exactly solvable: $\sigma^{-2} = (2/N)(1+U N/t)^{1/2}$. The single particle density matrix for these states reads
\begin{equation}
\rho_{\textsc{sp}} = \frac{N}{2} \left( \begin{matrix} 1&e^{-1/(2 \sigma^2)} \\ e^{-1/(2 \sigma^2)}&1 \end{matrix} \right).
\label{matrixfocktocoherent}
\end{equation}
It has eigenvalues $ \frac{N}{2} (1 \pm e^{-1/\sigma^2})$ and the fluctuation in the number of particles on each well is $\expval{(\Delta n_i)^2}=\sigma^2 /2$. As the gaussian width increases, the number of particles on each well vary from $(N,0)$ to $(N/2,N/2)$ and the number fluctuations $\expval{(\Delta n_i)^2}$ varies from $N$ to $0$. 

Let us discuss now the case of attractive interactions $U<0$. When the interaction is attractive, states having a huge difference in the number of particles on each well, or, in other words, a huge amount of particles on a single well, are favored. Thus, the effect of slowly turning on an atractive interaction is to split the Gaussian of the noninteracting state~\eqref{Gaussiandistrib} into a symmetric distribution with two peaks, a process which is captured by the family of states 
\begin{equation}
\Psi_{\ell}(a) = K \left( e^{-(\ell-a)^2/2 \sigma'^2} +  e^{-(\ell+a)^2/2 \sigma'^2} \right),
\end{equation}
being $2a$ is the separation between the peaks, $\sigma'$ their width and $K$ is a normalization factor. As $a$ varies from $0$ to $N/2$ and $\sigma'$ reduces at the same time from $1 / \sqrt{N}$ to 0, the coherent state of the free theory evolves to the cat state (which is reached in the $U/t \rightarrow -\infty$ limit).

\subsection{Effective spacetimes from BEC's}
\label{Subsection:Effective_SpacetimesE}

We discussed above that BECs are good analogue systems. Let us dig in the properties that our toy model needs to fullfill in order for the excitations propagating on top of it to obey a Klein-Gordon equation on a curved geometry. In order to build an analogue gravity model, a crucial necessary condition is that the system can exhibit in some regime \textit{causality properties}. For the Bose-Einstein condensate, this means that it should be able to accommodate the propagation of sound-like excitations within the system, which correspond to some of the collective excitations of the system. A fundamental ingredient for the propagation of this collective excitations is the existence of an effective repulsion among the constituents of the BEC. In addition to this condition, we require the additional condition that the wavelengths of the excitations are much greater than the healing length of the condensate, which roughly speaking is the length at which inhomogeneities of the atomic density composing the condensate are smoothened out, as we discussed above. 

The coherent state resulting from taking $U=0$ or the cat state that results from taking $U<0$ will not give rise to sound excitations in principle since the condensate is in a regime in which no repulsive interactions are present. Hence, it seems that it is not possible that any quantum sound emerges. Let us focus on the cat state, since it is the state that would give rise to an analogue of a superposition of spacetimes, as long as it admitted sound excitations. One could build initially a Schr\"{o}dinger-cat-type state from Eq.~\eqref{cat_state} displaying no sound-like excitations initially. For example, one could adiabatically switch on the attractive interactions ($U<0$) beginning with the $U=0$ case through Feshbach resonances. Once the system settles in the appropriate cat state, one could turn the interaction towards the repulsive regime in such a way that sound excitations emerge. We will examine the validity of this picture in the next section through a stability analysis.

\subsection{Stability of the ground states}
\label{Subsection:Stability}

We saw that even though one-particle density matrices of the Fock and cat states were the same, they described completely different states: while the Schr\"{o}dinger-cat state had enormous number fluctuations, the Fock state had zero number fluctuations; that is, higher order correlation functions are needed to characterize these different fragmented states. 

Moreover, these huge number fluctuations in the Schr\"{o}dinger-cat state are telling us something about the stability of the system: it is obvious that such a delocalized system must tend to be unstable under local perturbations which act independently on each of the wells. In fact, let us consider a generic interaction modelled by the Hamiltonian 
\begin{equation}
H' = \epsilon \left( c_{1}^{\dagger} a_1 + c_1 a_1^{\dagger} \right),
\label{perturbation}
\end{equation} 
where $ \epsilon$ is a small parameter controlling the perturbation, $a_1$ is the usual annihilation operator for bosons in the first well and $c_{1}^{\dagger}$ is the creation operator for a generic excited state of the first well and, consequently, representing a local perturbation of the first well. Under this small perturbation, the Fock state~\eqref{fock_state} is robust since it simply changes to another Fock state $\ket{F'}$ under the action of this Hamiltonian operator
\begin{equation}
\ket{F'} = H' \ket{F} = K' c_{1}^{\dagger} a_1^{\dagger (N/2-1)} a_2^{\dagger N/2} \ket{0},
\end{equation}
with $K'$ an irrelevant normalization constant, which shows the robustness of Fock states under standard local perturbations. However, the cat state does not display such robustness. Actually, the cat state tends to collapse to a localized condensate once we take them into account
\begin{equation}
H' \ket{\textrm{cat}} = K'' c_{1}^{\dagger} a_1^{\dagger (N-1)} \ket{0},
\end{equation}
with $K''$ another irrelevant normalization constant, since this state corresponds to a localized state on the first well. It corresponds to a number state with \mbox{$N-1$} particles on the ground state and one particle in the excited state described by the operator $c_{1}^{\dagger}$. 

The conclusion we extract from this analysis is that the system tends to avoid being in delocalized macroscopic superpositions since they are unstable under local perturbations that do not coherently hit the system on the two places. We have just considered interactions of the type~\eqref{perturbation} which will be contained in a generic condensed matter Hamiltonian. This can be seen by noting that, in general, we can work in the second quantized formalism in which the fundamental object we consider is the field $\psi(x)$,
\begin{equation}
H_{\textrm{int}} = \int d^3 x h(\psi ^{\dagger} (x), \psi(x)),
\end{equation}
whose expansion in a concrete basis of one-particle states reads
\begin{equation}
\psi = \sum_i \left( a_i f_i + a_i^{\dagger} f_i^{*} \right). 
\end{equation}
The previous simplification was equivalent to saying that we just focused on the mode $f_1$ which was an excited state of the first well. In practice we will have a tower of excited states above them that can be regarded as perturbations of the form~\eqref{perturbation}. 

\subsection{Attempting to generate a superposition of two effective spacetimes}
\label{Subsection:Superposing_Spacetimes}

Now that we have described the three main states that can appear in a Bose-Hubbard model, and their stability properties, it is time to discuss them at the light of their ability to provide superpositions of effective spacetimes. 

Our aim here is to consider a model whose substratum is an intrinsic quantum system, admitting an effective geometrical description in some regime, and analyze the implications and viability of producing a state which would be understood within the effective geometrical description as a putative superposition or mixture of spacetimes. For that purpose, we first need to identify the states that would correspond to superpositions of spacetimes. We have two natural candidates for this purpose: the coherent and the cat states. From an effective spacetime perspective, the Fock state, being a product state, would correspond to an incoherent mixture of two spacetimes. This means that the propagation of sound on each of the two wells would occur independently of what is happening on the other well.

The coherent state can be taken as representing a system in a macroscopically delocalized state. However, we have seen that this state appears at the cost of eliminating the local interactions between the composing bosons. Having negligible local interactions, this state lacks the possibility of producing a rich causal physics. For instance, for $U=0$ the state develops no acoustic excitations, it does not really serves as an analogue model of gravity. However, given our analysis of the previous section in which we see that there is a smooth interpolation among the states with $U=0$ and $U > 0$, we can turn on a small repulsion without breaking the qualitative features of the state and allowing sound-like excitations. However, this delocalized state would have a wave-function with a profile varying too fast spatially to allow for the hydrodynamic approximation to be accurate. So we see that the coherent state does not seem suitable to describe superpositions of spacetimes.

We are left with the cat state as the only one that we can identify, at least momentarily, with a superposition of two different spacetimes, since it is a superposition of two states representing an effectively curved geometry on each of the wells. Actually, it seems to be the naive state that one would also build in a quantum gravity theory as a putative superposition of spacetimes. 

Hence, the following question appears now: if we construct a condensate in the cat state, i.e., what we would claim to be a would-be ``superposition of spacetimes'' in the analogue system, what happens with the potential acoustic perturbations propagating on top of such a spacetime: is there an effective spacetime resulting from this superposition? The idea is that there should be a limitation in describing the acoustic perturbations propagating on top of such a superposition as acoustic perturbations on a new spacetime. Otherwise, this would mean that the superposition of analogue spacetimes would be also a spacetime, contrary to what one would expect to happen even with real gravitational systems. Indeed, such limitation appears in the form of a huge instability of the system under consideration. As we discussed in Subsection~\ref{Subsection:Stability}, the cat state tends to decay to a Fock-like state under small perturbations. The propagation of sound-like excitations itself would be enough to destabilize such superposition. This translates into a decay of the putative superpositon of spacetimes into a mixture of two spacetimes, i.e., two effective spacetimes which behave independently from each other on each of the wells. 

Apart from the instability under perturbations of this state which we have already discussed, an additional impediment appears. If we consider the evolution of such a system according to the Gross-Pitaevskii equation which describes generic fluid condensates, the interference is such that the healing length of this condensate is much bigger than the healing length of either of the condensates. Consequently, the effective spacetime picture breaks at much bigger wavelengths and we conclude that the hydrodynamic picture is, at best, hard to maintain for these superpositions. This manifests the impossibility of finding an analogue of a superposition of spacetimes. 

Actually, this impossibility recalls Penrose's ideas on the possibility of superposing spacetimes~\cite{Penrose2014,Penrose1992,Penrose1996}. On the one hand, Penrose argues that the superposition of spacetimes is an ill-posed concept in gravity, since it leads to a causal structure which is not well-defined. Since already in gravity there are hints that these superpositions are ill-posed, we could have expected in advance a fundamental limitation to find it in an analogue version. On the other hand, Penrose relates this ill-definitness of the causality of superposed spacetimes with the mechanisms of quantum state reduction. We will come back to Penrose's idea in Section~\ref{Section:Penrose_Idea}, where we will contrast it with our findings in this analogue setup.

\subsection{Same localization, different causality}
\label{Subsection:Same_Localization}

We will now discuss the viability of generating a superposition of spacetimes, but now we will consider a different situation. Instead of superposing two spacetimes with the same causal properties but based around different locations, we will now consider superposing two spacetimes at the same location with different causal properties. Roughly speaking, instead of attempting to superpose two geometries generated by two lumps of matter spatially translated, we want to superpose two lumps of matter with different shapes and densities located around the same region. The double well Bose-Hubbard model does not allow us to consider this situation and it is hard to design a toy model that allows to consider it. However, we will try to analyze in general terms whether this is at all possible or not. 

As the simplest situation, one could think of a homogeneous BEC at rest but with a superposition of sound velocities. We recall that the sound velocity in a condensate, $c_s^2=g n_c /m$, is controlled by the coupling constant (from the $\abs{\psi}^4$ term of the Gross-Pitaevskii equation) $g$, the effective mass of the bosons $m$ and their number density $n_c$. The numbers $g$ and $m$ are parameters of the system which are in principle not subject to quantum rules. The masses are renormalized due to the interactions and, in principle, it is even possible to achieve negative masses with suitable microscopic structure of the system~\cite{Khamehchi2017}. On the other hand, the coupling constant can be controlled by an external magnetic field by using Feshbach resonances~\cite{Chin2010}. Then, we can ask, whether we can engineer this magnetic field to be in a quantum superposition. The magnetic field we speak about is a macroscopic entity so what we are doing with this inquiry is just translating the problem of generating a macroscopic superposition from one place to another. Instead, it is typically assumed that the values of the parameter of a system are phenomenological values to which one could in principle associate infinitely precise values. As the causality of the system depends directly on these values one could think that they imprint a classical flavor into the notion of causality. Hence, this tuning of magnetic fields does not seem to be a good way of achieving such a superposition. 

The other quantity entering the definition of the sound velocity is the density of condensed particles $n_c$. This quantity can have fluctuations if we deal with a system allowing flows of particles (grand-canonical ensembles), but in a condensed matter system the number of particles is completely fixed: the creation of new particles is forbidden by a huge energy gap. If one were hypothetically able to generate a BEC in an ultrarelativistic regime, the possibility of having a superposition of states with different number of particles would exist. But again, would it be possible to have a stable superposition of two mutually non-interacting gases each having a different number of particles? The possibility does not sound realistic, the very notion of macroscopic (thermodynamic) behaviour would try to produce states with a peaked distribution for the number of particles. Actually, the standard ensemble in terms of which condensed matter systems are described is the Grand Canonical Ensemble, which allows not only for energy fluxes but also particle density fluxes between the system and the environment. In such scenarios, thermal homogeneity is needed to reach thermodynamical equilibrium although it is not sufficient. Chemical homogeneity is also required, which means having equal chemical potentials for the bath and the system. If the number of particles is not highly peaked, it is impossible to reach a stationary state. Hence, it seems that this is also a no-go pathway, since an almost stationary situation in which the condensate is equilibrated is required for the hydrodynamic picture to be sensible at all. 
       
The only element of the causality in a BEC remaining is the flow velocity of the fluid. The flow velocity that appears in the effective analogue gravity metric in a BEC corresponds to a macroscopic occupation of a particular phase structure for the mono-particle wave functions. Once again, the natural question is whether it is possible to have a stable superposition of condensates with different phases or not. The situation would be parallel to our discussion of the stability properties of the cat and Fock states, respectively. Depending on whether we perform a superposition of the two states with all the particles on each phase state or a more democratic Fock-like state in which half of the particles are on each of the phase states, we would produce an unstable state or a stable state, respectively. The additional difference is the external potential which, for the formation of standard condensates, always favors one of the two phase states. Thus, although under the absence of external potentials it would be possible to generate a stable Fock-like state, in a real systems the particles would always tend to be projected onto one of the two fixed phases states. 

Thus, this qualitative discussion suggests also the inability to perform superpositions of different causalities, i.e. to have a BEC sitting on a single location but giving rise to different causal structures for the propagation of sound-like waves. Nonetheless, further study is required to fully understand and characterize whether there is a universal constraint on the possibility of simulating such superposition of analogue geometries in arbitrary analogue systems. 

\section{Penrose's ideas on superposing spacetimes}
\label{Section:Penrose_Idea}

The results that we have found in our analogue system strongly resonate with Penrose's ideas about the \emph{instabilities} that one finds when superposing spacetimes~\cite{Penrose1989,Penrose1992,Penrose1996,Penrose2014}. Although it has been a minoritarian activity, there has been a steady flux of works trying to see whether the standard formalism and interpretation of quantum mechanics, which one could associate to von Neumann's formalization~\cite{vonneumann1955}, could fail in some regime~\cite{Wigner1962,Bialyanicki1976,Pearle1976,Ellis1983,Pearle1989,Ghirardi1986,Ghirardi1990,Bassi2013,Feldmann2012,Bassi2016}. Within those attempts, there have been many similar in spirit to Penrose's, i.e., attempts trying to trace these modifications of quantum mechanics back to a gravitational origin~\cite{Karolyhazy1966,Komar1969,Karolyhazy1974,Diosi1986,Coleman1988,Diosi1989,Gisin1989,Penrose1989,Ghirardi1990b,Penrose1992,Percival1995,Pearle1996,Penrose1996,Frenkel2002,Hu2003,Giulini2011,Adler2014,Hu2014,Penrose2014,Sharma2014,Bera2015,Singh2015,Donadi2020}. Here we are going to focus our comparison with Penrose's proposal as it is presented in the closest language to our analogue model. 

Penrose argues that superposing spacetimes is an ill-defined concept because the resulting state does not incorporate a clear definition of time translation. In turn, he argues that this uncertainty in the definition of time-translation should lead to an uncertainty in the definition of energy.  Furthermore, this should entail the decay of such states into mixed states where correlations among the superposed spacetimes are lost. His ideas are not at the level of a precise dynamical theory and hence we do not attempt to make a precise analogy here: we simply discuss its similarities and differences with our analogue toy model.

\begin{figure}
    \begin{center}
        \includegraphics[width = 0.22 \textwidth]{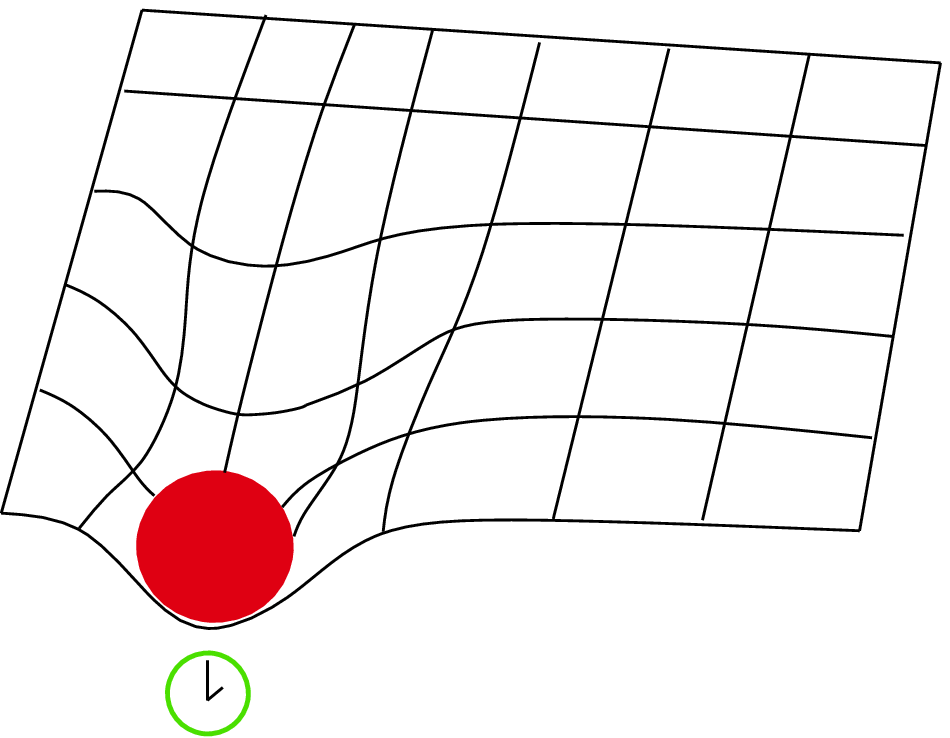}
        \includegraphics[width = 0.20 \textwidth]{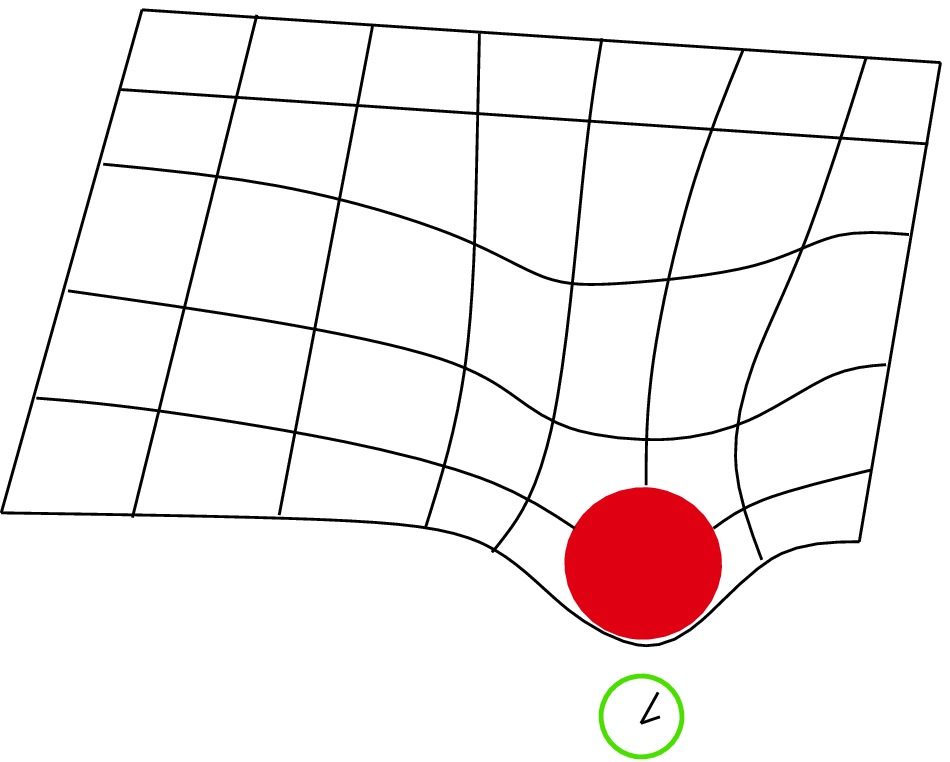}
        \caption{We pictorially represent the two configurations that we are want to superpose. They correspond to the gravitational field associated with a certain lump of matter (a star, for instance), which is located in two different places for each of the states.}
        \label{Fig:Penrose_Picture}
    \end{center}
\end{figure}

In the case in which we have a superposition of configurations based on different locations, for example, two spacetimes corresponding to a lump of matter (for example a star) located at two different positions. If the star is not too compact, the effective spacetimes can be well-described within the Newtonian limit.
For this setup, which is pictorially depicted in Fig.~\ref{Fig:Penrose_Picture}, Penrose is able to quantify the incompatibility of the superpositions. He defines a quantity $\Delta$ which is basically the gravitational self-energy of the difference between the mass distributions of each of the two locations of the lump of mass: 
\begin{equation}
    \Delta = - 4 \pi G \int d^3 x d^3 y \frac{\left[ \rho(x) - \rho'(x) \right] \left[ \rho(y) - \rho'(y) \right]}{ \abs{x-y}},
\end{equation}
where $\rho, \rho'$ are the mass densities of the two lumps. This quantity would control the lifetime $\tau$ of the superposition for these simple examples of cat-like states for the matter distribution. 

For a careful analysis of the physical impact of this mechanism of gravitationally induced decoherence and a comparison with the more standard enviromental decoherence~\cite{Zeh1970,Zurek1981,Zurek1982}, see~\cite{Howl2019}. Moreover, recently it has been argued that natural models with $\tau = \eta  \hbar /\Delta$ and $\eta$ of order 1 are falsified experimentally\footnote{We restore here $\hbar$ to match the formulas in the references.}: the relaxation times these models imply appear to be much shorter than the ones inferred in the experiment~\cite{Donadi2020}. However, it is important to recall that the indirect results of this experiment crucially rely on assuming a random diffusive behaviour for the density matrix of the system. For us it is not clear to what extent this needs to be a compulsory assumption.

The mechanism causing decoherence in our model, although similar in spirit to Penrose's proposal differs from it in several points. In our toy model the instability under local perturbations that the cat state suffers comes from the huge fluctuations in the particle number density that the system displays. Recalling that the speed of sound at a point in the condensate $c_s$ is proportional to the local number density of particles $n_c$, huge fluctuations in the number of particles would induce huge fluctuations of the underlying causal structure for the propagation of sound waves in the effective geometry. Such fluctuations arise because the cat state is highly delocalized. However, Penrose argues that the source of instability for the superpositions of spacetimes arises due to a purely geometric reason: it arises from the inability to make sense of the notion of stationarity for such states. The absence of a well-defined notion of time, i.e. the absence of a well-defined parameter with respect to which we can talk about evolution when considering these kind of superpositions makes impossible to regard the system as having some kind of causal structure. Although some fuzziness of the causal structure when entering the quantum regime is to be expected, attempting to superpose two radically different causalities would drive us completely into this regime where the causal structure itself dilutes and becomes a meaningless concept. Thus, the reason for the decay of a delocalized superposition seems to be the huge instability these states develop due to being far from having a well-defined causal structure. 

All mechanisms that try to account for the reduction of state in quantum mechanics need to introduce parameters controlling how macroscopic a system is. Standard environmentally induced decoherence generically uses the number of particles in the system $N$ as a measure of how macroscopic the system is. Penrose's model measures how complex and how macroscopic a system is in terms of the gravitational field it can create. In general terms, we can expect that for ordinary macroscopic matter both measures display similar behaviours regarding the decoherence time-scales, but conceptual differences appear. If one were able to monitor the whole set of particles composing a system, it would be possible in principle to distinguish among both mechanisms~\cite{Howl2019}.  

An important difference of our model with respect to Penrose's idea is that in BECs we can make sense of the superposition of two Minkoswki spacetimes with different speeds of sound. In Penrose's geometrical language these spacetimes are one and the same. If causality were an emergent property, much like the sound speed of collective excitations, then it would have additional properties beyond those associated with the effective geometrical description. Our model suggest that the inability of superposing different causalities is a robust idea that could survive in different approaches to quantum gravity. Although in a completely different setup, similar resonant conclusions are reached in Euclidean Dynamical Triangulations and Causal Dynamical Triangulations. In the former, one sums over arbitrary Euclidean manifolds and the result is dominated by configurations that do not resemble at all a smooth, almost flat on suitable scales, spacetime that we live in. On the other hand, in the latter one just sums over geometries that lead to a well-defined causal structure, hence constraining the geometries one sums over. In this case, it is possible to find almost macroscopic spacetimes~\cite{Loll2019}.

Something quite similar happens when one tries to use analogue models for constructing effective spacetimes with causal pathological behaviours, for instance, with chronological horizons~\cite{Barcelo2022}. In the general relativistic description, the presence of causal pathologies appears to be tied up just to the presence of exotic matter: the semiclassical regime does not uncover further limitations. However, playing with analogue models suggests that one might find further fundamental limitations, like the existence of an underlying fundamental causality, making the difficulty of violating causality principles also a robust idea.

One additional difference between the two mechanisms is that, strictly speaking, in our formalism the loss of coherence is only effective as the underlying system is a $N$-body quantum mechanical system. On the contrary, in Penrose's description, the gravitational decoherence appears as something fundamental. In a sense, our proposal seems to be closer to the environmental decoherence proposal than Penrose's. However, we rather think that our analysis highlights that the line separating an effective decoherence from a fundamental decoherence is blurred. Our toy model suggests that even taking a purely quantum theory as the high-energy theory, any practical and effective physical theory constructed by macroscopic observers would contain quantum elements but also some elements that should not be treated quantum mechanically: there are elements in the effective theory that are collective excitations of already quantized degrees of freedom and they must not be quantized again. The difference with standard environmental decoherence is that the environmental degrees of freedom in this case are not so controllable from the effective theory point of view.

As a final comment, let us discuss the interplay of our model and the problems that quantum mechanics faces, since they were Penrose's original motivation to propose his ideas. Quantum mechanics, as described by just a 
dynamical wavefunction, faces two related problems. The first one is how to accommodate the apparent absence of quantum superpositions of macroscopic systems. As the formalism \textit{per se} does not suggest a division between microscopic and macroscopic systems, the same rules for superposition should apply to both realms. The second problem, is how to account for the specific although probabilistic results of actual acts of measurement in the laboratories, i.e., how to account for the apparent intrinsically statistical result of individual acts of measurement. Without the additional Born rule, the unitary evolution of quantum mechanics does not even describes what is a measurement nor what is the source of non-determinism. As a disclaimer, let us recall here that a hidden-variable take on quantum mechanics such as that of Bohm provides a solution to these problems at the cost of introducing some explicit non-locality in its formulation~\cite{Bohm1952,Bohm1995}. As explained in the introduction, our discussion in this paper is focused on interpretations of quantum mechanics giving some reality just to the wavefunction.

Our model, like Penrose's proposal, directly accounts for the first of these problems: how a superposition ends up becoming a mixture (a state lacking quantum coherences). For the second problem, it does not offer a complete solution since it would require making an explicit model of measurement. Actually, this is a long-standing problem that has accumulated a vast literature~\cite{Schlosshauer2005}. Nonetheless, we feel that our analogue toy model adds some interesting new perspectives to this second conundrum.

The philosophy underlying the analogue gravity program and in a more complete sense, emergent gravity scenarios (i.e. hypothetical scenarios in which Einstein equations are also emergent~\cite{Barcelo2021}) is that everything that we perceive is constituted by low-energy collective excitations of some fundamental microscopic degrees of freedom. For instance, the spacetime itself and the quantum fields describing the Standard Model of Particle Physics would be part of such collective excitations. In that sense, although the microscopic theory may not display a lorentzian causality, the effective low-energy regime does. Within this logic, observers would also be ``emergent" in the sense of being composed of excitations of these effective fields. 
The notion of observer that we are using is that of a conscious being, understood not literally as a human being but as any primitive form of consciousness or even a recording device. In general any sufficiently complex system relying on causality does the job. In fact a crucial prerequisite for consciousness to appear is causality itself\footnote{There even exist works suggesting that our consciousness developed because of our progressive ability to simulate progressions of hypothetical events organized as causes and effects in our minds~\cite{Jordan2012}.}. 

Now, recall that the main conclusion that we have drawn up from our analysis is that causality is a special macroscopic property in the sense that it is not amenable to be in quantum superposition. Then, an immediate consequence of this tight relation between causality and consciousness seems to be that different states of consciousness will also tend to rapidly decohere. 

Our construction in this paper, takes standard non-relativistic quantum mechanics, call it NRQM1 as the microscopic fundamental theory with no special commitment to any of its interpretations. Then, inner observers living in the system could make another non-relativistic quantum theory (NRQM2) to describe the phenomena they perceive whenever the regimes they are probing are non-relativistic. NRQM2 should of course become a Lorentz-invariant quantum field theory at high energies from the internal perspective but still low energies from the perspective of laboratory observers, that would still use NRQM1. Now, consider for example that NRQM1 as a high-energy theory happens to provide only practical decoherence and nothing similar to a collapse to one of the reduced states. Even in this case the description of NRQM2 will tend to incorporate in its rules a collapse postulate. This will be so because conscious beings will never have fuzzy experiences associated with being in cat-like states, i.e. they will never perceived macroscopic quantum coherences between states with very different causalities. If on the contrary the formulation of NRQM1 incorporated a mechanism for the selection of one and only one alternative under individual acts of measurement (by incorporating for instance a stochastic rule, for example \emph{\`a la} Ghirardi-Rimini-Weber~\cite{Ghirardi1986}), the NRQM2 would have the same form as before. Thus, the phenomenology of NRQM2 is independent of the measurement theory taken in NRQM1. 
In this way our example illustrates how difficult it is to learn things about a microscopic fundamental theory if we only have access to the collective excitations belonging to its low-energy sector.  

This is closely related to Wigner's discussion on the role of consciousness in our physical description~\cite{Wigner1962}. Essentially, he advocated that sufficiently macroscopic or complex processes (not just those involving conscious beings as complex as human beings) will in practice decohere, marking a separation between the classical and quantum realms. Notice that here, complexity and macroscopic are used in a non-technical sense referring to an imprecise characterization of such magnitudes. The difference of our approach is that it highlights the decoherent character of causality as a more primitive notion than consciousness.

\section{Summary and Discussion}
\label{Section:Discussion}

Our aim in this work has been to present an analogue model of gravity where it is possible to analyze the properties of the states corresponding to would-be superposition of spacetimes. Although they are logically possible in theories of quantum gravity, we find difficulties when we try to engineer them in analogue gravity set-ups.

We have studied a toy model of a Bose-Einstein condensate in a double well which in a suitable regime gives rise to what we could call a superposition of spacetimes. This toy model shows three phases: a coherent phase, when interactions are negligible; the fragmented uncorrelated phase, whose representative in the regime in which the hopping dominates is the Fock state; and the fragmented correlated phase, whose representative when the interactions dominate is the so-called cat state. In principle, the coherent state cannot lead to causal behaviour of signals propagating on top of it. Furthermore, even if we turn on some repulsive interactions for the system to display such behaviour, the resulting state does not resemble at all what we would expect that a superposition of spacetimes looks like. The Fock state lacks quantum coherences and we regard it as a mixture of spacetimes, instead of a quantum superposition. It is stable under local perturbations and hence offers no difficulty either from the condensed matter perspective or from the point of view of its analogue gravity interpretation. The cat state, which we would interpret as an analogue of the superposition of spacetimes, is highly unstable under local perturbations, that tend to drive it to a Fock-like state. 

This is exactly the same phenomenology advocated by Penrose. Superposed spacetimes should rapidly decay into mixtures of spacetimes. The rate at which this decay occurs is proportional to a measure of how macroscopic the system is. Penrose measure is the magnitude of the gravitational field that the system is able to generate. In our model, this measure is given by the number of condensed particles and the ratio between the hopping parameter and the interactions ($t/U$). Penrose proposal is based on the fact that a superposition of two spacetimes is not a well-posed concept, since it does not exhibit any sensible causal structure. This happens even if one superposes stationary causal structures. In our model, the fluctuations in the number of particles on each well is what quantifies the instability of the BEC under local perturbations. We recall that the sound speed is proportional to the local number density of bosons, and, thus, huge fluctuations in the number density are to be interpreted as a huge undefiniteness of the effective causal structure of the system. In that sense, we find that our analogue model illustrates the difficulty in pushing the causal structure of analogue systems to its limits: systems avoid causal pathologies. This is similar to our previous findings that analogue systems avoid giving rise to chronologically ill-behaved spacetimes~\cite{Barcelo2022}. To put it explictly again, we have found that analogue systems tend to avoid configurations exhibiting superpositions of analogue spacetimes (or causalities); this means, configurations that do not have a well-defined causality. 

Thus, in tune with Penrose's ideas, the lesson we seem to be learning is that, although macroscopic superpositions might be relevant for very high energies (this situation is of course logically possible), there are arguments to suspect that they might be qualitatively irrelevant at low energies at which gravity could be described by an effective geometry. 

\begin{acknowledgements}
GGM would like to thank Alfredo Luis for very useful discussions. Financial support was provided by the Spanish Government through the projects PID2020-118159GB-C43, PID2020-118159GB-C44, FIS2017-86497-C2-1-P, FIS2017-86497-C2-2-P (with FEDER contribution), and by the Junta de Andaluc\'{\i}a through the project FQM219. CB and GGM acknowledge financial support from the State Agency for Research of the Spanish MCIU through the ``Center of Excellence Severo Ochoa" award to the Instituto de Astrof\'{\i}sica de Andaluc\'{\i}a (SEV-2017-0709). GGM is funded by the Spanish Government fellowship FPU20/01684. 
\end{acknowledgements}

\bibliography{penrose_biblio}

\begin{thebibliography}{65}%
\makeatletter
\providecommand \@ifxundefined [1]{%
 \@ifx{#1\undefined}
}%
\providecommand \@ifnum [1]{%
 \ifnum #1\expandafter \@firstoftwo
 \else \expandafter \@secondoftwo
 \fi
}%
\providecommand \@ifx [1]{%
 \ifx #1\expandafter \@firstoftwo
 \else \expandafter \@secondoftwo
 \fi
}%
\providecommand \natexlab [1]{#1}%
\providecommand \enquote  [1]{``#1''}%
\providecommand \bibnamefont  [1]{#1}%
\providecommand \bibfnamefont [1]{#1}%
\providecommand \citenamefont [1]{#1}%
\providecommand \href@noop [0]{\@secondoftwo}%
\providecommand \href [0]{\begingroup \@sanitize@url \@href}%
\providecommand \@href[1]{\@@startlink{#1}\@@href}%
\providecommand \@@href[1]{\endgroup#1\@@endlink}%
\providecommand \@sanitize@url [0]{\catcode `\\12\catcode `\$12\catcode
  `\&12\catcode `\#12\catcode `\^12\catcode `\_12\catcode `\%12\relax}%
\providecommand \@@startlink[1]{}%
\providecommand \@@endlink[0]{}%
\providecommand \url  [0]{\begingroup\@sanitize@url \@url }%
\providecommand \@url [1]{\endgroup\@href {#1}{\urlprefix }}%
\providecommand \urlprefix  [0]{URL }%
\providecommand \Eprint [0]{\href }%
\providecommand \doibase [0]{http://dx.doi.org/}%
\providecommand \selectlanguage [0]{\@gobble}%
\providecommand \bibinfo  [0]{\@secondoftwo}%
\providecommand \bibfield  [0]{\@secondoftwo}%
\providecommand \translation [1]{[#1]}%
\providecommand \BibitemOpen [0]{}%
\providecommand \bibitemStop [0]{}%
\providecommand \bibitemNoStop [0]{.\EOS\space}%
\providecommand \EOS [0]{\spacefactor3000\relax}%
\providecommand \BibitemShut  [1]{\csname bibitem#1\endcsname}%
\let\auto@bib@innerbib\@empty
\bibitem [{\citenamefont {Green}\ \emph
  {et~al.}(2012{\natexlab{a}})\citenamefont {Green}, \citenamefont {Schwarz},\
  and\ \citenamefont {Witten}}]{Green1987a}%
  \BibitemOpen
  \bibfield  {author} {\bibinfo {author} {\bibfnamefont {M.~B.}\ \bibnamefont
  {Green}}, \bibinfo {author} {\bibfnamefont {J.~H.}\ \bibnamefont {Schwarz}},
  \ and\ \bibinfo {author} {\bibfnamefont {E.}~\bibnamefont {Witten}},\ }\href
  {\doibase 10.1017/CBO9781139248563} {\emph {\bibinfo {title} {Superstring
  Theory: 25th Anniversary Edition}}},\ \bibinfo {series} {Cambridge Monographs
  on Mathematical Physics}, Vol.~\bibinfo {volume} {1}\ (\bibinfo  {publisher}
  {Cambridge University Press},\ \bibinfo {year} {2012})\BibitemShut {NoStop}%
\bibitem [{\citenamefont {Green}\ \emph
  {et~al.}(2012{\natexlab{b}})\citenamefont {Green}, \citenamefont {Schwarz},\
  and\ \citenamefont {Witten}}]{Green1987b}%
  \BibitemOpen
  \bibfield  {author} {\bibinfo {author} {\bibfnamefont {M.~B.}\ \bibnamefont
  {Green}}, \bibinfo {author} {\bibfnamefont {J.~H.}\ \bibnamefont {Schwarz}},
  \ and\ \bibinfo {author} {\bibfnamefont {E.}~\bibnamefont {Witten}},\ }\href
  {\doibase 10.1017/CBO9781139248563} {\emph {\bibinfo {title} {Superstring
  Theory: 25th Anniversary Edition}}},\ \bibinfo {series} {Cambridge Monographs
  on Mathematical Physics}, Vol.~\bibinfo {volume} {2}\ (\bibinfo  {publisher}
  {Cambridge University Press},\ \bibinfo {year} {2012})\BibitemShut {NoStop}%
\bibitem [{\citenamefont {Wheeler}(1987)}]{Wheeler1968}%
  \BibitemOpen
  \bibfield  {author} {\bibinfo {author} {\bibfnamefont {J.~A.}\ \bibnamefont
  {Wheeler}},\ }\href@noop {} {\bibfield  {journal} {\bibinfo  {journal} {Adv.
  Ser. Astrophys. Cosmol.}\ }\textbf {\bibinfo {volume} {3}},\ \bibinfo {pages}
  {27} (\bibinfo {year} {1987})}\BibitemShut {NoStop}%
\bibitem [{\citenamefont {Thiemann}(2007)}]{Thiemann2007}%
  \BibitemOpen
  \bibfield  {author} {\bibinfo {author} {\bibfnamefont {T.}~\bibnamefont
  {Thiemann}},\ }\href {\doibase 10.1017/CBO9780511755682} {\emph {\bibinfo
  {title} {{Modern Canonical Quantum General Relativity}}}},\ Cambridge
  Monographs on Mathematical Physics\ (\bibinfo  {publisher} {Cambridge
  University Press},\ \bibinfo {year} {2007})\BibitemShut {NoStop}%
\bibitem [{\citenamefont {Percacci}(2017)}]{Percacci2017}%
  \BibitemOpen
  \bibfield  {author} {\bibinfo {author} {\bibfnamefont {R.}~\bibnamefont
  {Percacci}},\ }\href {\doibase 10.1142/10369} {\emph {\bibinfo {title} {{An
  Introduction to Covariant Quantum Gravity and Asymptotic Safety}}}},\
  \bibinfo {series} {100 Years of General Relativity}, Vol.~\bibinfo {volume}
  {3}\ (\bibinfo  {publisher} {World Scientific},\ \bibinfo {year}
  {2017})\BibitemShut {NoStop}%
\bibitem [{\citenamefont {Bonanno}\ \emph {et~al.}(2020)\citenamefont
  {Bonanno}, \citenamefont {Eichhorn}, \citenamefont {Gies}, \citenamefont
  {Pawlowski}, \citenamefont {Percacci}, \citenamefont {Reuter}, \citenamefont
  {Saueressig},\ and\ \citenamefont {Vacca}}]{Bonanno2020}%
  \BibitemOpen
  \bibfield  {author} {\bibinfo {author} {\bibfnamefont {A.}~\bibnamefont
  {Bonanno}}, \bibinfo {author} {\bibfnamefont {A.}~\bibnamefont {Eichhorn}},
  \bibinfo {author} {\bibfnamefont {H.}~\bibnamefont {Gies}}, \bibinfo {author}
  {\bibfnamefont {J.~M.}\ \bibnamefont {Pawlowski}}, \bibinfo {author}
  {\bibfnamefont {R.}~\bibnamefont {Percacci}}, \bibinfo {author}
  {\bibfnamefont {M.}~\bibnamefont {Reuter}}, \bibinfo {author} {\bibfnamefont
  {F.}~\bibnamefont {Saueressig}}, \ and\ \bibinfo {author} {\bibfnamefont
  {G.~P.}\ \bibnamefont {Vacca}},\ }\href {\doibase 10.3389/fphy.2020.00269}
  {\bibfield  {journal} {\bibinfo  {journal} {Front. in Phys.}\ }\textbf
  {\bibinfo {volume} {8}},\ \bibinfo {pages} {269} (\bibinfo {year} {2020})},\
  \Eprint {http://arxiv.org/abs/2004.06810} {arXiv:2004.06810 [gr-qc]}
  \BibitemShut {NoStop}%
\bibitem [{\citenamefont {Loll}(2020)}]{Loll2019}%
  \BibitemOpen
  \bibfield  {author} {\bibinfo {author} {\bibfnamefont {R.}~\bibnamefont
  {Loll}},\ }\href {\doibase 10.1088/1361-6382/ab57c7} {\bibfield  {journal}
  {\bibinfo  {journal} {Class. Quant. Grav.}\ }\textbf {\bibinfo {volume}
  {37}},\ \bibinfo {pages} {013002} (\bibinfo {year} {2020})},\ \Eprint
  {http://arxiv.org/abs/1905.08669} {arXiv:1905.08669 [hep-th]} \BibitemShut
  {NoStop}%
\bibitem [{\citenamefont {Surya}(2019)}]{Surya2019}%
  \BibitemOpen
  \bibfield  {author} {\bibinfo {author} {\bibfnamefont {S.}~\bibnamefont
  {Surya}},\ }\href {\doibase 10.1007/s41114-019-0023-1} {\bibfield  {journal}
  {\bibinfo  {journal} {Living Rev. Rel.}\ }\textbf {\bibinfo {volume} {22}},\
  \bibinfo {pages} {5} (\bibinfo {year} {2019})},\ \Eprint
  {http://arxiv.org/abs/1903.11544} {arXiv:1903.11544 [gr-qc]} \BibitemShut
  {NoStop}%
\bibitem [{\citenamefont {Brune}\ \emph {et~al.}(1992)\citenamefont {Brune},
  \citenamefont {Haroche}, \citenamefont {Raimond}, \citenamefont
  {Davidovich},\ and\ \citenamefont {Zagury}}]{Brune1992}%
  \BibitemOpen
  \bibfield  {author} {\bibinfo {author} {\bibfnamefont {M.}~\bibnamefont
  {Brune}}, \bibinfo {author} {\bibfnamefont {S.}~\bibnamefont {Haroche}},
  \bibinfo {author} {\bibfnamefont {J.~M.}\ \bibnamefont {Raimond}}, \bibinfo
  {author} {\bibfnamefont {L.}~\bibnamefont {Davidovich}}, \ and\ \bibinfo
  {author} {\bibfnamefont {N.}~\bibnamefont {Zagury}},\ }\href {\doibase
  10.1103/PhysRevA.45.5193} {\bibfield  {journal} {\bibinfo  {journal} {Phys.
  Rev. A}\ }\textbf {\bibinfo {volume} {45}},\ \bibinfo {pages} {5193}
  (\bibinfo {year} {1992})}\BibitemShut {NoStop}%
\bibitem [{\citenamefont {Barcelo}\ \emph {et~al.}(2005)\citenamefont
  {Barcelo}, \citenamefont {Liberati},\ and\ \citenamefont
  {Visser}}]{Barcelo2011}%
  \BibitemOpen
  \bibfield  {author} {\bibinfo {author} {\bibfnamefont {C.}~\bibnamefont
  {Barcelo}}, \bibinfo {author} {\bibfnamefont {S.}~\bibnamefont {Liberati}}, \
  and\ \bibinfo {author} {\bibfnamefont {M.}~\bibnamefont {Visser}},\ }\href
  {\doibase 10.12942/lrr-2005-12} {\bibfield  {journal} {\bibinfo  {journal}
  {Living Rev. Rel.}\ }\textbf {\bibinfo {volume} {8}},\ \bibinfo {pages} {12}
  (\bibinfo {year} {2005})},\ \Eprint {http://arxiv.org/abs/gr-qc/0505065}
  {arXiv:gr-qc/0505065} \BibitemShut {NoStop}%
\bibitem [{\citenamefont {Wald}(1984)}]{Wald1984}%
  \BibitemOpen
  \bibfield  {author} {\bibinfo {author} {\bibfnamefont {R.~M.}\ \bibnamefont
  {Wald}},\ }\href {\doibase 10.7208/chicago/9780226870373.001.0001} {\emph
  {\bibinfo {title} {{General Relativity}}}}\ (\bibinfo  {publisher} {Chicago
  Univ. Pr.},\ \bibinfo {address} {Chicago, USA},\ \bibinfo {year}
  {1984})\BibitemShut {NoStop}%
\bibitem [{\citenamefont {Garay}(1999)}]{Garay1999}%
  \BibitemOpen
  \bibfield  {author} {\bibinfo {author} {\bibfnamefont {L.~J.}\ \bibnamefont
  {Garay}},\ }\href {\doibase 10.1142/S0217751X99001913} {\bibfield  {journal}
  {\bibinfo  {journal} {International Journal of Modern Physics A}\ }\textbf
  {\bibinfo {volume} {14}},\ \bibinfo {pages} {4079} (\bibinfo {year}
  {1999})},\ \Eprint
  {http://arxiv.org/abs/https://doi.org/10.1142/S0217751X99001913}
  {https://doi.org/10.1142/S0217751X99001913} \BibitemShut {NoStop}%
\bibitem [{\citenamefont {Steinhauer}(2016)}]{Steinhauer2016}%
  \BibitemOpen
  \bibfield  {author} {\bibinfo {author} {\bibfnamefont {J.}~\bibnamefont
  {Steinhauer}},\ }\href {\doibase 10.1038/nphys3863} {\bibfield  {journal}
  {\bibinfo  {journal} {Nature Physics}\ }\textbf {\bibinfo {volume} {12}},\
  \bibinfo {pages} {959} (\bibinfo {year} {2016})}\BibitemShut {NoStop}%
\bibitem [{\citenamefont {Visser}(1998)}]{Visser1997}%
  \BibitemOpen
  \bibfield  {author} {\bibinfo {author} {\bibfnamefont {M.}~\bibnamefont
  {Visser}},\ }\href {\doibase 10.1088/0264-9381/15/6/024} {\bibfield
  {journal} {\bibinfo  {journal} {Class. Quant. Grav.}\ }\textbf {\bibinfo
  {volume} {15}},\ \bibinfo {pages} {1767} (\bibinfo {year} {1998})},\ \Eprint
  {http://arxiv.org/abs/gr-qc/9712010} {arXiv:gr-qc/9712010} \BibitemShut
  {NoStop}%
\bibitem [{\citenamefont {Penrose}(2014)}]{Penrose2014}%
  \BibitemOpen
  \bibfield  {author} {\bibinfo {author} {\bibfnamefont {R.}~\bibnamefont
  {Penrose}},\ }\href {\doibase 10.1007/s10701-013-9770-0} {\bibfield
  {journal} {\bibinfo  {journal} {Found. Phys.}\ }\textbf {\bibinfo {volume}
  {44}},\ \bibinfo {pages} {557} (\bibinfo {year} {2014})}\BibitemShut
  {NoStop}%
\bibitem [{\citenamefont {Zeh}(1970)}]{Zeh1970}%
  \BibitemOpen
  \bibfield  {author} {\bibinfo {author} {\bibfnamefont {H.~D.}\ \bibnamefont
  {Zeh}},\ }\href {\doibase 10.1007/BF00708656} {\bibfield  {journal} {\bibinfo
   {journal} {Foundations of Physics}\ }\textbf {\bibinfo {volume} {1}},\
  \bibinfo {pages} {69} (\bibinfo {year} {1970})}\BibitemShut {NoStop}%
\bibitem [{\citenamefont {Zurek}(1981)}]{Zurek1981}%
  \BibitemOpen
  \bibfield  {author} {\bibinfo {author} {\bibfnamefont {W.~H.}\ \bibnamefont
  {Zurek}},\ }\href {\doibase 10.1103/PhysRevD.24.1516} {\bibfield  {journal}
  {\bibinfo  {journal} {Phys. Rev. D}\ }\textbf {\bibinfo {volume} {24}},\
  \bibinfo {pages} {1516} (\bibinfo {year} {1981})}\BibitemShut {NoStop}%
\bibitem [{\citenamefont {Zurek}(1982)}]{Zurek1982}%
  \BibitemOpen
  \bibfield  {author} {\bibinfo {author} {\bibfnamefont {W.~H.}\ \bibnamefont
  {Zurek}},\ }\href {\doibase 10.1103/PhysRevD.26.1862} {\bibfield  {journal}
  {\bibinfo  {journal} {Phys. Rev. D}\ }\textbf {\bibinfo {volume} {26}},\
  \bibinfo {pages} {1862} (\bibinfo {year} {1982})}\BibitemShut {NoStop}%
\bibitem [{\citenamefont {Penrose}(2016)}]{Penrose1989}%
  \BibitemOpen
  \bibfield  {author} {\bibinfo {author} {\bibfnamefont {R.}~\bibnamefont
  {Penrose}},\ }\href {https://books.google.es/books?id=X28sDwAAQBAJ} {\emph
  {\bibinfo {title} {The Emperor's New Mind: Concerning Computers, Minds, and
  the Laws of Physics}}},\ Oxford landmark science\ (\bibinfo  {publisher}
  {Oxford University Press},\ \bibinfo {year} {2016})\BibitemShut {NoStop}%
\bibitem [{\citenamefont {Penrose}(1992)}]{Penrose1992}%
  \BibitemOpen
  \bibfield  {author} {\bibinfo {author} {\bibfnamefont {R.}~\bibnamefont
  {Penrose}},\ }in\ \href@noop {} {\emph {\bibinfo {booktitle} {{13th
  Conference on General Relativity and Gravitation (GR-13)}}}}\ (\bibinfo
  {year} {1992})\BibitemShut {NoStop}%
\bibitem [{\citenamefont {Penrose}(1996)}]{Penrose1996}%
  \BibitemOpen
  \bibfield  {author} {\bibinfo {author} {\bibfnamefont {R.}~\bibnamefont
  {Penrose}},\ }\href {\doibase 10.1007/BF02105068} {\bibfield  {journal}
  {\bibinfo  {journal} {Gen. Rel. Grav.}\ }\textbf {\bibinfo {volume} {28}},\
  \bibinfo {pages} {581} (\bibinfo {year} {1996})}\BibitemShut {NoStop}%
\bibitem [{\citenamefont {Barcel\'o}\ \emph {et~al.}(2022)\citenamefont
  {Barcel\'o}, \citenamefont {S\'anchez}, \citenamefont {Garc\'\i{}a-Moreno},\
  and\ \citenamefont {Jannes}}]{Barcelo2022}%
  \BibitemOpen
  \bibfield  {author} {\bibinfo {author} {\bibfnamefont {C.}~\bibnamefont
  {Barcel\'o}}, \bibinfo {author} {\bibfnamefont {J.~E.}\ \bibnamefont
  {S\'anchez}}, \bibinfo {author} {\bibfnamefont {G.}~\bibnamefont
  {Garc\'\i{}a-Moreno}}, \ and\ \bibinfo {author} {\bibfnamefont
  {G.}~\bibnamefont {Jannes}},\ }\href {\doibase
  10.1140/epjc/s10052-022-10275-3} {\bibfield  {journal} {\bibinfo  {journal}
  {Eur. Phys. J. C}\ }\textbf {\bibinfo {volume} {82}},\ \bibinfo {pages} {299}
  (\bibinfo {year} {2022})},\ \Eprint {http://arxiv.org/abs/2201.11072}
  {arXiv:2201.11072 [gr-qc]} \BibitemShut {NoStop}%
\bibitem [{\citenamefont {Hawking}\ and\ \citenamefont
  {Ellis}(2011)}]{Hawking1973}%
  \BibitemOpen
  \bibfield  {author} {\bibinfo {author} {\bibfnamefont {S.~W.}\ \bibnamefont
  {Hawking}}\ and\ \bibinfo {author} {\bibfnamefont {G.~F.~R.}\ \bibnamefont
  {Ellis}},\ }\href {\doibase 10.1017/CBO9780511524646} {\emph {\bibinfo
  {title} {{The Large Scale Structure of Space-Time}}}},\ Cambridge Monographs
  on Mathematical Physics\ (\bibinfo  {publisher} {Cambridge University
  Press},\ \bibinfo {year} {2011})\BibitemShut {NoStop}%
\bibitem [{\citenamefont {Garay}\ \emph {et~al.}(2001)\citenamefont {Garay},
  \citenamefont {Anglin}, \citenamefont {Cirac},\ and\ \citenamefont
  {Zoller}}]{Garay2000}%
  \BibitemOpen
  \bibfield  {author} {\bibinfo {author} {\bibfnamefont {L.~J.}\ \bibnamefont
  {Garay}}, \bibinfo {author} {\bibfnamefont {J.~R.}\ \bibnamefont {Anglin}},
  \bibinfo {author} {\bibfnamefont {J.~I.}\ \bibnamefont {Cirac}}, \ and\
  \bibinfo {author} {\bibfnamefont {P.}~\bibnamefont {Zoller}},\ }\href
  {\doibase 10.1103/PhysRevA.63.023611} {\bibfield  {journal} {\bibinfo
  {journal} {Phys. Rev. A}\ }\textbf {\bibinfo {volume} {63}},\ \bibinfo
  {pages} {023611} (\bibinfo {year} {2001})},\ \Eprint
  {http://arxiv.org/abs/gr-qc/0005131} {arXiv:gr-qc/0005131} \BibitemShut
  {NoStop}%
\bibitem [{\citenamefont {Lewenstein}\ \emph {et~al.}(2012)\citenamefont
  {Lewenstein}, \citenamefont {Sanpera},\ and\ \citenamefont
  {Ahufinger}}]{Lewenstein2012}%
  \BibitemOpen
  \bibfield  {author} {\bibinfo {author} {\bibfnamefont {M.}~\bibnamefont
  {Lewenstein}}, \bibinfo {author} {\bibfnamefont {A.}~\bibnamefont {Sanpera}},
  \ and\ \bibinfo {author} {\bibfnamefont {V.}~\bibnamefont {Ahufinger}},\
  }\href {https://books.google.it/books?id=Wpl91RDxV5IC} {\emph {\bibinfo
  {title} {Ultracold Atoms in Optical Lattices: Simulating quantum many-body
  systems}}}\ (\bibinfo  {publisher} {OUP Oxford},\ \bibinfo {year}
  {2012})\BibitemShut {NoStop}%
\bibitem [{\citenamefont {Mueller}\ \emph {et~al.}(2006)\citenamefont
  {Mueller}, \citenamefont {Ho}, \citenamefont {Ueda},\ and\ \citenamefont
  {Baym}}]{Mueller2006}%
  \BibitemOpen
  \bibfield  {author} {\bibinfo {author} {\bibfnamefont {E.~J.}\ \bibnamefont
  {Mueller}}, \bibinfo {author} {\bibfnamefont {T.}~\bibnamefont {Ho}},
  \bibinfo {author} {\bibfnamefont {M.}~\bibnamefont {Ueda}}, \ and\ \bibinfo
  {author} {\bibfnamefont {G.}~\bibnamefont {Baym}},\ }\href {\doibase
  10.1103/PhysRevA.74.033612} {\bibfield  {journal} {\bibinfo  {journal} {Phys.
  Rev. A}\ }\textbf {\bibinfo {volume} {74}},\ \bibinfo {pages} {033612}
  (\bibinfo {year} {2006})}\BibitemShut {NoStop}%
\bibitem [{\citenamefont {Leggett}(2006)}]{Leggett2006}%
  \BibitemOpen
  \bibfield  {author} {\bibinfo {author} {\bibfnamefont {A.}~\bibnamefont
  {Leggett}},\ }\href {https://books.google.es/books?id=PiRRAAAAMAAJ} {\emph
  {\bibinfo {title} {Quantum Liquids: Bose Condensation and Cooper Pairing in
  Condensed-matter Systems}}},\ Oxford Graduate Texts\ (\bibinfo  {publisher}
  {OUP Oxford},\ \bibinfo {year} {2006})\BibitemShut {NoStop}%
\bibitem [{\citenamefont {Khamehchi}\ \emph {et~al.}(2017)\citenamefont
  {Khamehchi}, \citenamefont {Hossain}, \citenamefont {Mossman}, \citenamefont
  {Zhang}, \citenamefont {Busch}, \citenamefont {Forbes},\ and\ \citenamefont
  {Engels}}]{Khamehchi2017}%
  \BibitemOpen
  \bibfield  {author} {\bibinfo {author} {\bibfnamefont {M.~A.}\ \bibnamefont
  {Khamehchi}}, \bibinfo {author} {\bibfnamefont {K.}~\bibnamefont {Hossain}},
  \bibinfo {author} {\bibfnamefont {M.~E.}\ \bibnamefont {Mossman}}, \bibinfo
  {author} {\bibfnamefont {Y.}~\bibnamefont {Zhang}}, \bibinfo {author}
  {\bibfnamefont {T.}~\bibnamefont {Busch}}, \bibinfo {author} {\bibfnamefont
  {M.~M.}\ \bibnamefont {Forbes}}, \ and\ \bibinfo {author} {\bibfnamefont
  {P.}~\bibnamefont {Engels}},\ }\href {\doibase
  10.1103/PhysRevLett.118.155301} {\bibfield  {journal} {\bibinfo  {journal}
  {Phys. Rev. Lett.}\ }\textbf {\bibinfo {volume} {118}},\ \bibinfo {pages}
  {155301} (\bibinfo {year} {2017})}\BibitemShut {NoStop}%
\bibitem [{\citenamefont {Chin}\ \emph {et~al.}(2010)\citenamefont {Chin},
  \citenamefont {Grimm}, \citenamefont {Julienne},\ and\ \citenamefont
  {Tiesinga}}]{Chin2010}%
  \BibitemOpen
  \bibfield  {author} {\bibinfo {author} {\bibfnamefont {C.}~\bibnamefont
  {Chin}}, \bibinfo {author} {\bibfnamefont {R.}~\bibnamefont {Grimm}},
  \bibinfo {author} {\bibfnamefont {P.}~\bibnamefont {Julienne}}, \ and\
  \bibinfo {author} {\bibfnamefont {E.}~\bibnamefont {Tiesinga}},\ }\href
  {\doibase 10.1103/RevModPhys.82.1225} {\bibfield  {journal} {\bibinfo
  {journal} {Rev. Mod. Phys.}\ }\textbf {\bibinfo {volume} {82}},\ \bibinfo
  {pages} {1225} (\bibinfo {year} {2010})}\BibitemShut {NoStop}%
\bibitem [{\citenamefont {von Neumann}\ and\ \citenamefont
  {Beyer}(1955)}]{vonneumann1955}%
  \BibitemOpen
  \bibfield  {author} {\bibinfo {author} {\bibfnamefont {J.}~\bibnamefont {von
  Neumann}}\ and\ \bibinfo {author} {\bibfnamefont {R.}~\bibnamefont {Beyer}},\
  }\href {https://books.google.es/books?id=JLyCo3RO4qUC} {\emph {\bibinfo
  {title} {Mathematical Foundations of Quantum Mechanics}}},\ Goldstine Printed
  Materials\ (\bibinfo  {publisher} {Princeton University Press},\ \bibinfo
  {year} {1955})\BibitemShut {NoStop}%
\bibitem [{\citenamefont {Wigner}(1995)}]{Wigner1962}%
  \BibitemOpen
  \bibfield  {author} {\bibinfo {author} {\bibfnamefont {E.~P.}\ \bibnamefont
  {Wigner}},\ }\enquote {\bibinfo {title} {Remarks on the mind-body
  question},}\ in\ \href {\doibase 10.1007/978-3-642-78374-6_20} {\emph
  {\bibinfo {booktitle} {Philosophical Reflections and Syntheses}}},\ \bibinfo
  {editor} {edited by\ \bibinfo {editor} {\bibfnamefont {J.}~\bibnamefont
  {Mehra}}}\ (\bibinfo  {publisher} {Springer Berlin Heidelberg},\ \bibinfo
  {address} {Berlin, Heidelberg},\ \bibinfo {year} {1995})\ p.\ \bibinfo
  {pages} {247}\BibitemShut {NoStop}%
\bibitem [{\citenamefont {Bialynicki-Birula}\ and\ \citenamefont
  {Mycielski}(1976)}]{Bialyanicki1976}%
  \BibitemOpen
  \bibfield  {author} {\bibinfo {author} {\bibfnamefont {I.}~\bibnamefont
  {Bialynicki-Birula}}\ and\ \bibinfo {author} {\bibfnamefont {J.}~\bibnamefont
  {Mycielski}},\ }\href {\doibase https://doi.org/10.1016/0003-4916(76)90057-9}
  {\bibfield  {journal} {\bibinfo  {journal} {Annals of Physics}\ }\textbf
  {\bibinfo {volume} {100}},\ \bibinfo {pages} {62} (\bibinfo {year}
  {1976})}\BibitemShut {NoStop}%
\bibitem [{\citenamefont {Pearle}(1976)}]{Pearle1976}%
  \BibitemOpen
  \bibfield  {author} {\bibinfo {author} {\bibfnamefont {P.}~\bibnamefont
  {Pearle}},\ }\href {\doibase 10.1103/PhysRevD.13.857} {\bibfield  {journal}
  {\bibinfo  {journal} {Phys. Rev. D}\ }\textbf {\bibinfo {volume} {13}},\
  \bibinfo {pages} {857} (\bibinfo {year} {1976})}\BibitemShut {NoStop}%
\bibitem [{\citenamefont {Ellis}\ \emph {et~al.}(1984)\citenamefont {Ellis},
  \citenamefont {Hagelin}, \citenamefont {Nanopoulos},\ and\ \citenamefont
  {Srednicki}}]{Ellis1983}%
  \BibitemOpen
  \bibfield  {author} {\bibinfo {author} {\bibfnamefont {J.}~\bibnamefont
  {Ellis}}, \bibinfo {author} {\bibfnamefont {J.~S.}\ \bibnamefont {Hagelin}},
  \bibinfo {author} {\bibfnamefont {D.}~\bibnamefont {Nanopoulos}}, \ and\
  \bibinfo {author} {\bibfnamefont {M.}~\bibnamefont {Srednicki}},\ }\href
  {\doibase https://doi.org/10.1016/0550-3213(84)90053-1} {\bibfield  {journal}
  {\bibinfo  {journal} {Nuclear Physics B}\ }\textbf {\bibinfo {volume}
  {241}},\ \bibinfo {pages} {381} (\bibinfo {year} {1984})}\BibitemShut
  {NoStop}%
\bibitem [{\citenamefont {Pearle}(1989)}]{Pearle1989}%
  \BibitemOpen
  \bibfield  {author} {\bibinfo {author} {\bibfnamefont {P.}~\bibnamefont
  {Pearle}},\ }\href {\doibase 10.1103/PhysRevA.39.2277} {\bibfield  {journal}
  {\bibinfo  {journal} {Phys. Rev. A}\ }\textbf {\bibinfo {volume} {39}},\
  \bibinfo {pages} {2277} (\bibinfo {year} {1989})}\BibitemShut {NoStop}%
\bibitem [{\citenamefont {Ghirardi}\ \emph {et~al.}(1986)\citenamefont
  {Ghirardi}, \citenamefont {Rimini},\ and\ \citenamefont
  {Weber}}]{Ghirardi1986}%
  \BibitemOpen
  \bibfield  {author} {\bibinfo {author} {\bibfnamefont {G.~C.}\ \bibnamefont
  {Ghirardi}}, \bibinfo {author} {\bibfnamefont {A.}~\bibnamefont {Rimini}}, \
  and\ \bibinfo {author} {\bibfnamefont {T.}~\bibnamefont {Weber}},\ }\href
  {\doibase 10.1103/PhysRevD.34.470} {\bibfield  {journal} {\bibinfo  {journal}
  {Phys. Rev. D}\ }\textbf {\bibinfo {volume} {34}},\ \bibinfo {pages} {470}
  (\bibinfo {year} {1986})}\BibitemShut {NoStop}%
\bibitem [{\citenamefont {Ghirardi}\ \emph
  {et~al.}(1990{\natexlab{a}})\citenamefont {Ghirardi}, \citenamefont
  {Pearle},\ and\ \citenamefont {Rimini}}]{Ghirardi1990}%
  \BibitemOpen
  \bibfield  {author} {\bibinfo {author} {\bibfnamefont {G.}~\bibnamefont
  {Ghirardi}}, \bibinfo {author} {\bibfnamefont {P.}~\bibnamefont {Pearle}}, \
  and\ \bibinfo {author} {\bibfnamefont {A.}~\bibnamefont {Rimini}},\ }\href
  {\doibase 10.1103/PhysRevA.42.78} {\bibfield  {journal} {\bibinfo  {journal}
  {Phys. Rev. A}\ }\textbf {\bibinfo {volume} {42}},\ \bibinfo {pages} {78}
  (\bibinfo {year} {1990}{\natexlab{a}})}\BibitemShut {NoStop}%
\bibitem [{\citenamefont {Bassi}\ \emph {et~al.}(2013)\citenamefont {Bassi},
  \citenamefont {Lochan}, \citenamefont {Satin}, \citenamefont {Singh},\ and\
  \citenamefont {Ulbricht}}]{Bassi2013}%
  \BibitemOpen
  \bibfield  {author} {\bibinfo {author} {\bibfnamefont {A.}~\bibnamefont
  {Bassi}}, \bibinfo {author} {\bibfnamefont {K.}~\bibnamefont {Lochan}},
  \bibinfo {author} {\bibfnamefont {S.}~\bibnamefont {Satin}}, \bibinfo
  {author} {\bibfnamefont {T.~P.}\ \bibnamefont {Singh}}, \ and\ \bibinfo
  {author} {\bibfnamefont {H.}~\bibnamefont {Ulbricht}},\ }\href {\doibase
  10.1103/RevModPhys.85.471} {\bibfield  {journal} {\bibinfo  {journal} {Rev.
  Mod. Phys.}\ }\textbf {\bibinfo {volume} {85}},\ \bibinfo {pages} {471}
  (\bibinfo {year} {2013})}\BibitemShut {NoStop}%
\bibitem [{\citenamefont {Feldmann}\ and\ \citenamefont
  {Tumulka}(2012)}]{Feldmann2012}%
  \BibitemOpen
  \bibfield  {author} {\bibinfo {author} {\bibfnamefont {W.}~\bibnamefont
  {Feldmann}}\ and\ \bibinfo {author} {\bibfnamefont {R.}~\bibnamefont
  {Tumulka}},\ }\href {\doibase 10.1088/1751-8113/45/6/065304} {\bibfield
  {journal} {\bibinfo  {journal} {Journal of Physics A: Mathematical and
  Theoretical}\ }\textbf {\bibinfo {volume} {45}},\ \bibinfo {pages} {065304}
  (\bibinfo {year} {2012})}\BibitemShut {NoStop}%
\bibitem [{\citenamefont {Bilardello}\ \emph {et~al.}(2016)\citenamefont
  {Bilardello}, \citenamefont {Donadi}, \citenamefont {Vinante},\ and\
  \citenamefont {Bassi}}]{Bassi2016}%
  \BibitemOpen
  \bibfield  {author} {\bibinfo {author} {\bibfnamefont {M.}~\bibnamefont
  {Bilardello}}, \bibinfo {author} {\bibfnamefont {S.}~\bibnamefont {Donadi}},
  \bibinfo {author} {\bibfnamefont {A.}~\bibnamefont {Vinante}}, \ and\
  \bibinfo {author} {\bibfnamefont {A.}~\bibnamefont {Bassi}},\ }\href
  {\doibase https://doi.org/10.1016/j.physa.2016.06.134} {\bibfield  {journal}
  {\bibinfo  {journal} {Physica A: Statistical Mechanics and its Applications}\
  }\textbf {\bibinfo {volume} {462}},\ \bibinfo {pages} {764} (\bibinfo {year}
  {2016})}\BibitemShut {NoStop}%
\bibitem [{\citenamefont {Karolyhazy}(1966)}]{Karolyhazy1966}%
  \BibitemOpen
  \bibfield  {author} {\bibinfo {author} {\bibfnamefont {F.}~\bibnamefont
  {Karolyhazy}},\ }\href {\doibase 10.1007/BF02717926} {\bibfield  {journal}
  {\bibinfo  {journal} {Il Nuovo Cimento A (1965-1970)}\ }\textbf {\bibinfo
  {volume} {42}},\ \bibinfo {pages} {390} (\bibinfo {year} {1966})}\BibitemShut
  {NoStop}%
\bibitem [{\citenamefont {Komar}(1969)}]{Komar1969}%
  \BibitemOpen
  \bibfield  {author} {\bibinfo {author} {\bibfnamefont {A.}~\bibnamefont
  {Komar}},\ }\href {\doibase 10.1007/BF00669563} {\bibfield  {journal}
  {\bibinfo  {journal} {International Journal of Theoretical Physics}\ }\textbf
  {\bibinfo {volume} {2}},\ \bibinfo {pages} {157} (\bibinfo {year}
  {1969})}\BibitemShut {NoStop}%
\bibitem [{\citenamefont {Karolyhazy}(1974)}]{Karolyhazy1974}%
  \BibitemOpen
  \bibfield  {author} {\bibinfo {author} {\bibfnamefont {F.}~\bibnamefont
  {Karolyhazy}},\ }\href {http://cds.cern.ch/record/430144} {\bibfield
  {journal} {\bibinfo  {journal} {Magyar Fizikai Folyoirat}\ }\textbf {\bibinfo
  {volume} {22}},\ \bibinfo {pages} {23} (\bibinfo {year} {1974})}\BibitemShut
  {NoStop}%
\bibitem [{\citenamefont {Dio\'si}(1987)}]{Diosi1986}%
  \BibitemOpen
  \bibfield  {author} {\bibinfo {author} {\bibfnamefont {L.}~\bibnamefont
  {Dio\'si}},\ }\href {\doibase https://doi.org/10.1016/0375-9601(87)90681-5}
  {\bibfield  {journal} {\bibinfo  {journal} {Physics Letters A}\ }\textbf
  {\bibinfo {volume} {120}},\ \bibinfo {pages} {377} (\bibinfo {year}
  {1987})}\BibitemShut {NoStop}%
\bibitem [{\citenamefont {Coleman}(1988)}]{Coleman1988}%
  \BibitemOpen
  \bibfield  {author} {\bibinfo {author} {\bibfnamefont {S.~R.}\ \bibnamefont
  {Coleman}},\ }\href {\doibase 10.1016/0550-3213(88)90110-1} {\bibfield
  {journal} {\bibinfo  {journal} {Nucl. Phys. B}\ }\textbf {\bibinfo {volume}
  {307}},\ \bibinfo {pages} {867} (\bibinfo {year} {1988})}\BibitemShut
  {NoStop}%
\bibitem [{\citenamefont {Di\'osi}(1989)}]{Diosi1989}%
  \BibitemOpen
  \bibfield  {author} {\bibinfo {author} {\bibfnamefont {L.}~\bibnamefont
  {Di\'osi}},\ }\href {\doibase 10.1103/PhysRevA.40.1165} {\bibfield  {journal}
  {\bibinfo  {journal} {Phys. Rev. A}\ }\textbf {\bibinfo {volume} {40}},\
  \bibinfo {pages} {1165} (\bibinfo {year} {1989})}\BibitemShut {NoStop}%
\bibitem [{\citenamefont {Gisin}(1989)}]{Gisin1989}%
  \BibitemOpen
  \bibfield  {author} {\bibinfo {author} {\bibfnamefont {N.}~\bibnamefont
  {Gisin}},\ }\href@noop {} {\bibfield  {journal} {\bibinfo  {journal} {Helv.
  Phys. Acta}\ }\textbf {\bibinfo {volume} {62}},\ \bibinfo {pages} {363}
  (\bibinfo {year} {1989})}\BibitemShut {NoStop}%
\bibitem [{\citenamefont {Ghirardi}\ \emph
  {et~al.}(1990{\natexlab{b}})\citenamefont {Ghirardi}, \citenamefont
  {Grassi},\ and\ \citenamefont {Rimini}}]{Ghirardi1990b}%
  \BibitemOpen
  \bibfield  {author} {\bibinfo {author} {\bibfnamefont {G.}~\bibnamefont
  {Ghirardi}}, \bibinfo {author} {\bibfnamefont {R.}~\bibnamefont {Grassi}}, \
  and\ \bibinfo {author} {\bibfnamefont {A.}~\bibnamefont {Rimini}},\ }\href
  {\doibase 10.1103/PhysRevA.42.1057} {\bibfield  {journal} {\bibinfo
  {journal} {Phys. Rev. A}\ }\textbf {\bibinfo {volume} {42}},\ \bibinfo
  {pages} {1057} (\bibinfo {year} {1990}{\natexlab{b}})}\BibitemShut {NoStop}%
\bibitem [{\citenamefont {Percival}(1995)}]{Percival1995}%
  \BibitemOpen
  \bibfield  {author} {\bibinfo {author} {\bibfnamefont {I.~C.}\ \bibnamefont
  {Percival}},\ }\href {\doibase 10.1098/rspa.1995.0139} {\bibfield  {journal}
  {\bibinfo  {journal} {Proc. Roy. Soc. Lond. A}\ }\textbf {\bibinfo {volume}
  {451}},\ \bibinfo {pages} {503} (\bibinfo {year} {1995})},\ \Eprint
  {http://arxiv.org/abs/quant-ph/9508021} {arXiv:quant-ph/9508021} \BibitemShut
  {NoStop}%
\bibitem [{\citenamefont {Pearle}\ and\ \citenamefont
  {Squires}(1996)}]{Pearle1996}%
  \BibitemOpen
  \bibfield  {author} {\bibinfo {author} {\bibfnamefont {P.}~\bibnamefont
  {Pearle}}\ and\ \bibinfo {author} {\bibfnamefont {E.}~\bibnamefont
  {Squires}},\ }\href {\doibase 10.1007/BF02069474} {\bibfield  {journal}
  {\bibinfo  {journal} {Foundations of Physics}\ }\textbf {\bibinfo {volume}
  {26}},\ \bibinfo {pages} {291} (\bibinfo {year} {1996})}\BibitemShut
  {NoStop}%
\bibitem [{\citenamefont {Frenkel}(2002)}]{Frenkel2002}%
  \BibitemOpen
  \bibfield  {author} {\bibinfo {author} {\bibfnamefont {A.}~\bibnamefont
  {Frenkel}},\ }\href {\doibase 10.1023/A:1016057026165} {\bibfield  {journal}
  {\bibinfo  {journal} {Foundations of Physics}\ }\textbf {\bibinfo {volume}
  {32}},\ \bibinfo {pages} {751} (\bibinfo {year} {2002})}\BibitemShut
  {NoStop}%
\bibitem [{\citenamefont {Hu}\ and\ \citenamefont {Verdaguer}(2008)}]{Hu2003}%
  \BibitemOpen
  \bibfield  {author} {\bibinfo {author} {\bibfnamefont {B.~L.}\ \bibnamefont
  {Hu}}\ and\ \bibinfo {author} {\bibfnamefont {E.}~\bibnamefont {Verdaguer}},\
  }\href {\doibase 10.12942/lrr-2008-3} {\bibfield  {journal} {\bibinfo
  {journal} {Living Rev. Rel.}\ }\textbf {\bibinfo {volume} {11}},\ \bibinfo
  {pages} {3} (\bibinfo {year} {2008})},\ \Eprint
  {http://arxiv.org/abs/0802.0658} {arXiv:0802.0658 [gr-qc]} \BibitemShut
  {NoStop}%
\bibitem [{\citenamefont {Giulini}\ and\ \citenamefont
  {Gro{\ss}ardt}(2011)}]{Giulini2011}%
  \BibitemOpen
  \bibfield  {author} {\bibinfo {author} {\bibfnamefont {D.}~\bibnamefont
  {Giulini}}\ and\ \bibinfo {author} {\bibfnamefont {A.}~\bibnamefont
  {Gro{\ss}ardt}},\ }\href {\doibase 10.1088/0264-9381/28/19/195026} {\bibfield
   {journal} {\bibinfo  {journal} {Classical and Quantum Gravity}\ }\textbf
  {\bibinfo {volume} {28}},\ \bibinfo {pages} {195026} (\bibinfo {year}
  {2011})}\BibitemShut {NoStop}%
\bibitem [{\citenamefont {{Adler}}(2014)}]{Adler2014}%
  \BibitemOpen
  \bibfield  {author} {\bibinfo {author} {\bibfnamefont {S.~L.}\ \bibnamefont
  {{Adler}}},\ }\href@noop {} {\bibfield  {journal} {\bibinfo  {journal} {arXiv
  e-prints}\ ,\ \bibinfo {eid} {arXiv:1401.0353}} (\bibinfo {year} {2014})},\
  \Eprint {http://arxiv.org/abs/1401.0353} {arXiv:1401.0353 [gr-qc]}
  \BibitemShut {NoStop}%
\bibitem [{\citenamefont {Hu}(2014)}]{Hu2014}%
  \BibitemOpen
  \bibfield  {author} {\bibinfo {author} {\bibfnamefont {B.~L.}\ \bibnamefont
  {Hu}},\ }\href {\doibase 10.1088/1742-6596/504/1/012021} {\bibfield
  {journal} {\bibinfo  {journal} {J. Phys. Conf. Ser.}\ }\textbf {\bibinfo
  {volume} {504}},\ \bibinfo {pages} {012021} (\bibinfo {year} {2014})},\
  \Eprint {http://arxiv.org/abs/1402.6584} {arXiv:1402.6584 [gr-qc]}
  \BibitemShut {NoStop}%
\bibitem [{\citenamefont {Sharma}\ and\ \citenamefont
  {Singh}(2014)}]{Sharma2014}%
  \BibitemOpen
  \bibfield  {author} {\bibinfo {author} {\bibfnamefont {A.}~\bibnamefont
  {Sharma}}\ and\ \bibinfo {author} {\bibfnamefont {T.~P.}\ \bibnamefont
  {Singh}},\ }\href {\doibase 10.1142/S0218271814420073} {\bibfield  {journal}
  {\bibinfo  {journal} {International Journal of Modern Physics D}\ }\textbf
  {\bibinfo {volume} {23}},\ \bibinfo {pages} {1442007} (\bibinfo {year}
  {2014})},\ \Eprint
  {http://arxiv.org/abs/https://doi.org/10.1142/S0218271814420073}
  {https://doi.org/10.1142/S0218271814420073} \BibitemShut {NoStop}%
\bibitem [{\citenamefont {Bera}\ \emph {et~al.}(2015)\citenamefont {Bera},
  \citenamefont {Donadi}, \citenamefont {Lochan},\ and\ \citenamefont
  {Singh}}]{Bera2015}%
  \BibitemOpen
  \bibfield  {author} {\bibinfo {author} {\bibfnamefont {S.}~\bibnamefont
  {Bera}}, \bibinfo {author} {\bibfnamefont {S.}~\bibnamefont {Donadi}},
  \bibinfo {author} {\bibfnamefont {K.}~\bibnamefont {Lochan}}, \ and\ \bibinfo
  {author} {\bibfnamefont {T.~P.}\ \bibnamefont {Singh}},\ }\href {\doibase
  10.1007/s10701-015-9933-2} {\bibfield  {journal} {\bibinfo  {journal}
  {Foundations of Physics}\ }\textbf {\bibinfo {volume} {45}},\ \bibinfo
  {pages} {1537} (\bibinfo {year} {2015})}\BibitemShut {NoStop}%
\bibitem [{\citenamefont {Singh}(2015)}]{Singh2015}%
  \BibitemOpen
  \bibfield  {author} {\bibinfo {author} {\bibfnamefont {T.~P.}\ \bibnamefont
  {Singh}},\ }\href {\doibase 10.1088/1742-6596/626/1/012009} {\bibfield
  {journal} {\bibinfo  {journal} {J. Phys. Conf. Ser.}\ }\textbf {\bibinfo
  {volume} {626}},\ \bibinfo {pages} {012009} (\bibinfo {year} {2015})},\
  \Eprint {http://arxiv.org/abs/1503.01040} {arXiv:1503.01040 [quant-ph]}
  \BibitemShut {NoStop}%
\bibitem [{\citenamefont {Donadi}\ \emph {et~al.}(2021)\citenamefont {Donadi},
  \citenamefont {Piscicchia}, \citenamefont {Curceanu}, \citenamefont
  {Di{\'o}si}, \citenamefont {Laubenstein},\ and\ \citenamefont
  {Bassi}}]{Donadi2020}%
  \BibitemOpen
  \bibfield  {author} {\bibinfo {author} {\bibfnamefont {S.}~\bibnamefont
  {Donadi}}, \bibinfo {author} {\bibfnamefont {K.}~\bibnamefont {Piscicchia}},
  \bibinfo {author} {\bibfnamefont {C.}~\bibnamefont {Curceanu}}, \bibinfo
  {author} {\bibfnamefont {L.}~\bibnamefont {Di{\'o}si}}, \bibinfo {author}
  {\bibfnamefont {M.}~\bibnamefont {Laubenstein}}, \ and\ \bibinfo {author}
  {\bibfnamefont {A.}~\bibnamefont {Bassi}},\ }\href {\doibase
  10.1038/s41567-020-1008-4} {\bibfield  {journal} {\bibinfo  {journal} {Nature
  Physics}\ }\textbf {\bibinfo {volume} {17}},\ \bibinfo {pages} {74} (\bibinfo
  {year} {2021})}\BibitemShut {NoStop}%
\bibitem [{\citenamefont {Howl}\ \emph {et~al.}(2019)\citenamefont {Howl},
  \citenamefont {Penrose},\ and\ \citenamefont {Fuentes}}]{Howl2019}%
  \BibitemOpen
  \bibfield  {author} {\bibinfo {author} {\bibfnamefont {R.}~\bibnamefont
  {Howl}}, \bibinfo {author} {\bibfnamefont {R.}~\bibnamefont {Penrose}}, \
  and\ \bibinfo {author} {\bibfnamefont {I.}~\bibnamefont {Fuentes}},\ }\href
  {\doibase 10.1088/1367-2630/ab104a} {\bibfield  {journal} {\bibinfo
  {journal} {New J. Phys.}\ }\textbf {\bibinfo {volume} {21}},\ \bibinfo
  {pages} {043047} (\bibinfo {year} {2019})},\ \Eprint
  {http://arxiv.org/abs/1812.04630} {arXiv:1812.04630 [quant-ph]} \BibitemShut
  {NoStop}%
\bibitem [{\citenamefont {Bohm}(1952)}]{Bohm1952}%
  \BibitemOpen
  \bibfield  {author} {\bibinfo {author} {\bibfnamefont {D.}~\bibnamefont
  {Bohm}},\ }\href {\doibase 10.1103/PhysRev.85.166} {\bibfield  {journal}
  {\bibinfo  {journal} {Phys. Rev.}\ }\textbf {\bibinfo {volume} {85}},\
  \bibinfo {pages} {166} (\bibinfo {year} {1952})}\BibitemShut {NoStop}%
\bibitem [{\citenamefont {Bohm}\ and\ \citenamefont {Hiley}(1993)}]{Bohm1995}%
  \BibitemOpen
  \bibfield  {author} {\bibinfo {author} {\bibfnamefont {D.}~\bibnamefont
  {Bohm}}\ and\ \bibinfo {author} {\bibfnamefont {B.}~\bibnamefont {Hiley}},\
  }\href {https://books.google.es/books?id=Erp0kUhm\_NwC} {\emph {\bibinfo
  {title} {The Undivided Universe: An Ontological Interpretation of Quantum
  Theory}}},\ Physics, philosophy\ (\bibinfo  {publisher} {Routledge},\
  \bibinfo {year} {1993})\BibitemShut {NoStop}%
\bibitem [{\citenamefont {{Schlosshauer}}(2004)}]{Schlosshauer2005}%
  \BibitemOpen
  \bibfield  {author} {\bibinfo {author} {\bibfnamefont {M.}~\bibnamefont
  {{Schlosshauer}}},\ }\href {\doibase 10.1103/RevModPhys.76.1267} {\bibfield
  {journal} {\bibinfo  {journal} {Reviews of Modern Physics}\ }\textbf
  {\bibinfo {volume} {76}},\ \bibinfo {pages} {1267} (\bibinfo {year}
  {2004})},\ \Eprint {http://arxiv.org/abs/quant-ph/0312059}
  {arXiv:quant-ph/0312059 [quant-ph]} \BibitemShut {NoStop}%
\bibitem [{\citenamefont {Barcel\'o}\ \emph {et~al.}(2021)\citenamefont
  {Barcel\'o}, \citenamefont {Carballo-Rubio}, \citenamefont {Garay},\ and\
  \citenamefont {Garc\'\i{}a-Moreno}}]{Barcelo2021}%
  \BibitemOpen
  \bibfield  {author} {\bibinfo {author} {\bibfnamefont {C.}~\bibnamefont
  {Barcel\'o}}, \bibinfo {author} {\bibfnamefont {R.}~\bibnamefont
  {Carballo-Rubio}}, \bibinfo {author} {\bibfnamefont {L.~J.}\ \bibnamefont
  {Garay}}, \ and\ \bibinfo {author} {\bibfnamefont {G.}~\bibnamefont
  {Garc\'\i{}a-Moreno}},\ }\href {\doibase 10.3390/app11188763} {\bibfield
  {journal} {\bibinfo  {journal} {Appl. Sciences}\ }\textbf {\bibinfo {volume}
  {11}},\ \bibinfo {pages} {8763} (\bibinfo {year} {2021})},\ \Eprint
  {http://arxiv.org/abs/2108.06582} {arXiv:2108.06582 [gr-qc]} \BibitemShut
  {NoStop}%
\bibitem [{\citenamefont {Jordan}(2012)}]{Jordan2012}%
  \BibitemOpen
  \bibfield  {author} {\bibinfo {author} {\bibfnamefont {J.~S.}\ \bibnamefont
  {Jordan}},\ }\enquote {\bibinfo {title} {Consciousness and embodiment},}\ in\
  \href {\doibase 10.1007/978-1-4419-0463-8_279} {\emph {\bibinfo {booktitle}
  {Encyclopedia of the History of Psychological Theories}}},\ \bibinfo {editor}
  {edited by\ \bibinfo {editor} {\bibfnamefont {R.~W.}\ \bibnamefont
  {Rieber}}}\ (\bibinfo  {publisher} {Springer US},\ \bibinfo {address} {New
  York, NY},\ \bibinfo {year} {2012})\ pp.\ \bibinfo {pages}
  {217--222}\BibitemShut {NoStop}%
\end{thebibliography}%

\end{document}